\documentclass[11pt]{article}

\input epsf
\usepackage{latexsym}
\usepackage{amssymb}
\usepackage{amsmath}
\usepackage[dvips]{graphicx}
\usepackage{array}

\def\d3{^{(3)}\nabla}

\begin{document}
\centerline{\Large \bf CMB anisotropies in the presence of a stochastic magnetic field}

\vskip 2 cm

\centerline{Kerstin E. Kunze
\footnote{E-mail: kkunze@usal.es} }

\vskip 0.3cm

\centerline{{\sl Departamento de F\'\i sica Fundamental} and {\sl IUFFyM},}
\centerline{{\sl Universidad de Salamanca,}}
\centerline{{\sl Plaza de la Merced s/n, E-37008 Salamanca, Spain }}

\vskip 1.5cm

\centerline{\bf Abstract}
\vskip 0.5cm
\noindent
Primordial magnetic fields present since before the epoch of matter-radiation equality
have an effect on the anisotropies of the cosmic microwave background. The 
CMB anisotropies due to scalar perturbations are calculated in the  gauge invariant formalism for  magnetized adiabatic initial conditions. 
Furthermore the linear matter power spectrum is calculated. Numerical solutions are complemented by a qualitative analysis.

\vskip 1cm

\section{Introduction}

Observations of the cosmic microwave background (CMB) provide an important tool for testing the understanding of  the physics of the early universe. 
Since the first data of the angular power spectrum of the temperature fluctuations in 
the CMB provided by COBE \cite{cobe} the aim has been to put limits on cosmological parameters 
such as its matter composition or present day expansion rate. Moreover, with more and better quality data not only of the temperature fluctuations but also the polarization of the CMB becoming available at a fast rate it is possible to test even earlier stages of the universe long before the
beginning of the standard radiation dominated era. Thus different models of inflation are under scrutiny and part of the class of  models has already been found to be excluded by observations
\cite{modinf}.

Apart from matter, radiation and dark energy there is evidence for
magnetic fields on small upto very large scales. 
Magnetic fields on galactic scales have a field strength of the order of a $\mu$G and correlation lengths of the order of few kpc. There are also observations of magnetic fields in clusters which have similar field strengths and correlation lengths. More recently there have been indications of the existence of magnetic fields in high redshift galaxies \cite{cosm}. 

There exist a multitude of proposals of generation mechanisms of magnetic fields on scales comparable to galactic scales. Generally speaking there are two types of generation mechanisms.
Firstly models where magnetic fields are generated in the very early universe during inflation
due to the amplification of perturbations in the electromagnetic field. 
Secondly battery-type models operating after inflation has finished \cite{cosm}.

In most models magnetic fields are present long before the initial conditions for the evolution of the 
metric and matter perturbations are set, significant for the formation of the characteristic spectrum of the anisotropies of the CMB and its polarization. Therefore taking into account the presence of such a primordial magnetic field provides an interesting tool to put limits on its characteristics such as its field strength or spectral index.
This  has been already studied in different settings. The effect of the primordial magnetic field on the anisotropies of the cosmic microwave background (CMB) depends on its 
nature.  Assuming the magnetic field to be stochastic guaranties that it does not break the global isotropy of space-time \cite{magstoc}. Moreover there have also been
investigations where the primordial magnetic field is assumed to be uniform, which for large values
of the field strength would lead to models incompatible with the observed large scale isotropy of the universe \cite{maguni}.
Here we are going to assume that the magnetic field is primordial in origin and thus present since long before matter-radiation equality and, moreover, it is a stochastic magnetic field, so  that global isotropy is preserved.

There are several approaches to formulate the perturbation equations depending on whether a particular gauge is chosen, such as for example the synchronous gauge or a covariant formulation which is based on the fluid description. In this latter approach no particular gauge is chosen.
The calculation of the anisotropies of the temperature and the polarization of the CMB requires to solve the Boltzmann hierarchy describing the evolution of the photon distribution function.
COSMICS \cite{cosmics} was the first numerical Boltzmann solver which was followed by CMBFAST \cite{cmbfast} which uses the synchronous gauge to describe the perturbation equations and introduced the line-of-sight integration speeding up significantly the numerical calculation of the CMB anisotropies. CAMB \cite{camb} is another numerical code to solve the Boltzmann hierarchy and determine the CMB anisotropies. It uses the covariant approach to formulate the perturbation equations. More recently CMBEASY \cite{cmbeasy} yet another Boltzmann solver has been released. It allows to calculate the CMB anisotropies using the gauge invariant formalism to formulate the perturbation equations. Apart from the different mathematical approaches used to formulate the perturbation equations it is interesting to note the evolution from the point of view of programming languages. Whereas COSMICS and CMBFAST are written in Fortran 77, CAMB uses Fortran 90/95 and CMBEASY is written as a completely object-oriented C++  programme.
The effect of a primordial magnetic field on the CMB anisotropies has been calculated using different approaches: synchronous gauge and thus  a modified version of CMBFAST in 
\cite{grant,GK}  or the covariant formalism and a modified version of CAMB in \cite{fin1,fin2,le}.

Here the gauge-invariant formulation of the perturbation equations is used and a modified version of CMBEASY to calculate the CMB anisotropies due to  the scalar perturbations of the geometry.

In section 2 the perturbation equations and the initial conditions in the gauge-invariant formalism are presented.
Section 3 is devoted to the calculation of the different magnetic field contributions which due to the stochastic nature of the magnetic field involves convolution integrals. The magnetic field spectrum is damped due to diffusion on small scales. This is implemented here by using  a gaussian window function. In section 4 the angular power spectra determining the temperature and polarization auto-and cross-correlation functions and the linear power matter spectrum are presented. Section 5 contains the conclusions.

\section{Gauge-invariant description}
\setcounter{equation}{0}

In the gauge-invariant formalism the perturbation equations are written in terms of gauge-invariant variables. For the scalar sector Einstein's equations in Fourier space are given by \cite{ks,dmsw}
\begin{eqnarray}
\Phi&=&\frac{a^2\bar{\rho}\Delta+3a^2\bar{\rho}(1+w){\cal H}k^{-1}V}
{2\bar{M}^2_{\rm P}k^2+3a^2(1+w)\bar{\rho}}\\
\Psi&=&-\Phi-\frac{a^2\bar{p}\Pi}{\bar{M}^2_{\rm P}k^2}\\
\dot{\Phi}&=&{\cal H}\Psi-\frac{a^2(\bar{\rho}+\bar{p})V}{2\bar{M}^2_{\rm P}k},
\end{eqnarray}
where $\Phi$ and $\Psi$ are the gauge-invariant Bardeen potentials \cite{bard}.
A dot indicates the derivative with respect to conformal time of the unperturbed metric of the background space-time, that is 
$ds^2=a^2(\tau)(-d\tau^2+\delta_{ij}dx^idx^j)$, and $a(\tau)$ is the corresponding scale factor.
$\Delta$ is the gauge invariant density perturbation defined by 
\begin{eqnarray}
\Delta=\delta+3(1+w)\left(\Phi+{\cal H}k^{-1}\sigma\right),
\label{delta}
\end{eqnarray}
where $\delta\equiv\frac{\rho-\bar{\rho}}{\bar{\rho}}$ is the coefficient of the  density contrast in the harmonic expansion and $\sigma$ is the coefficient of the shear in the harmonic expansion. 
Photons ($\gamma$), neutrinos ($\nu$), baryons (b), cold dark matter (c) and the magnetic field (B) contribute to the total energy density perturbation, so that 
\begin{eqnarray}
\bar{\rho}\Delta&=&\rho_{\gamma}(\Delta_{\gamma}+\Delta_{\rm B})+\rho_{\nu}\Delta_{\nu}+\rho_{\rm c}\Delta_{\rm c}+\rho_{\rm b}\Delta_{\rm b}.
\end{eqnarray}
The total contribution to the gauge-invariant velocity $V$ \cite{ks,dmsw} and the total anisotropic stress are given by
\begin{eqnarray}
(1+w)\bar{\rho}V&=&\frac{4}{3}(\rho_{\gamma}V_{\gamma}+\rho_{\nu}V_{\nu})+\rho_{\rm c}V_{\rm c}+\rho_{\rm b}V_{\rm b},\\
\bar{p}\Pi&=&\frac{1}{3}\rho_{\gamma}(\pi_{\gamma}+\pi_{\rm B})+\frac{1}{3}\rho_{\nu}\pi_{\nu}.
\end{eqnarray}
Finally, 
\begin{eqnarray}
(1+w)\bar{\rho}=\frac{4}{3}(\rho_{\gamma}+\rho_{\nu})+\rho_{\rm b}+\rho_{\rm c}.
\end{eqnarray}
Furthermore $\bar{M}_{\rm P}\equiv M_{\rm P}/\sqrt{8\pi}$ is the reduced Planck mass and 
${\cal H}=\frac{\dot{a}}{a}$ leading to the form  of the Friedmann equation,
\begin{eqnarray}
{\cal H}^2=\frac{a^2}{3\bar{M}_{\rm P}^2}\rho.
\end{eqnarray}
Well  within the radiation dominated epoch electrons and baryons, coupled by Coulomb interaction, are as well tightly coupled to the photons because of Thomson scattering of the photons off the free electrons. Thus the baryon-electron-photon system is very well approximated by a one-fluid description. Due to the presence of the magnetic field this baryon-electron-photon fluid is magnetized. As shown in \cite{MG1} the system is in the magnetohydrodynamic limit.
The baryons are subject to the Lorentz force which changes the evolution of the baryon velocity in Fourier space to (cf., e.g., \cite{MG1,kr,fin1,fin2,le})
\begin{eqnarray}
\dot{V}_{\rm b}=(3c_s^2-1){\cal H}V_{\rm b}+k(\Psi-3c_s^2\Phi)+kc_s^2\Delta_{\rm b}+R\tau_c^{-1}(V_{\gamma}-V_{\rm b})+\frac{R}{4}kL,
\label{vb}
\end{eqnarray}
where $L$ is due to the Lorentz force $\vec{J}\times\vec{B}$ and 
$R\equiv\frac{4}{3}\frac{\rho_{\gamma}}{\rho_{\rm b}}$. Furthermore, $c_s^2=\frac{\partial\bar{p}}{\partial\bar{\rho}}$ is the adiabatic sound speed and $\tau_c^{-1}$ is the mean free path of photons between scatterings given in terms of the number density of free electrons $n_{\rm e}$ and the Thomson cross section $\sigma_{\rm T}$, $\tau_c^{-1}=an_{\rm e}\sigma_{\rm T}$.
In section \ref{sec3} the contribution of the magnetic field given by its energy density, anisotropic stress and Lorentz term will be specified.

\subsection{The tight-coupling limit}

In the very early stages, long before recombination, the energy density of free electrons scales as 
$a^{-3}$ so that the time between scatterings of photons is proportional to $a^2$. Thus the mean free time between scatterings of the photons is much smaller than the Hubble time which leads to a comparatively large value of $\tau_c^{-1}$. This implies that in the numerical  integration of the baryon and the photon velocities special care has to be taken with respect to the time step size.
This problem was solved by using an iterative solution during the very early stages in the tight coupling limit and afterwards the original equations \cite{py,mb,dor}.
To derive the equations for the evolution of the photon and the baryon velocities in the tight coupling limit we start by giving the relevant 
equations determining the photon evolution \cite{ks,dmsw},
\begin{eqnarray}
\dot{\Delta}_{\gamma}&=&-\frac{4}{3}kV_{\gamma}
\label{dg}\\
\dot{V}_{\gamma}&=&k(\Psi-\Phi)+\frac{k}{4}\Delta_{\gamma}-\frac{k}{6}\pi_{\gamma}+\tau_c^{-1}(V_{\rm b}-V_{\gamma})
\end{eqnarray}
and the baryon density contrast evolves as,
\begin{eqnarray}
\dot{\Delta}_{\rm b}&=&-kV_{\rm b}-3c_s^2{\cal H}\Delta_{\rm b}.
\end{eqnarray}
The baryon velocity is determined by equation (\ref{vb}).
From these equations using that $R\sim a^{-1}$ and $c_s^2\sim a^{-1}$ the so-called slip equation is derived $\dot{\cal V}\equiv\dot{V}_{\rm b}-\dot{V}_{\gamma}$ \cite{dor} which is 
found to be,
\begin{eqnarray}
\dot{\cal V}&=&\left[1+2\frac{\dot{a}}{a}\frac{\tau_c}{1+R}\right]^{-1}\left[\frac{\tau_c}{1+R}\left[-\frac{\ddot{a}}{a}V_{\rm b}+\ddot{V}_{\gamma}-\ddot{V}_{\rm b}-2\frac{\cal H}{\tau_c}\left(V_{\rm b}-V_{\gamma}\right)-{\cal H}k\left(\Psi-2\Phi+\frac{\Delta_{\gamma}}{2}-\frac{\pi_{\gamma}}{3}\right)
\right.\right.\nonumber\\
&&\left.\left.
+k\left(c_s^2\dot{\Delta}_{\rm b}-\frac{\dot{\Delta}_{\gamma}}{4}+\dot{\Phi}+\frac{\dot{\pi}_{\gamma}}{6}\right)\right]+\frac{\dot{\tau}_c}{\tau_c}\left(V_{\rm b}-V_{\gamma}\right)\right].
\end{eqnarray}
In the tight coupling limit the term $\ddot{V}_{\rm b}-\ddot{V}_{\gamma}$ is neglected \cite{dor}. 
In terms of $\dot{\cal V}$ the equations for the photon and baryon velocities are given by,
\begin{eqnarray}
\dot{V}_{\gamma}&=&\frac{R}{1+R}k\left(\frac{\Delta_{\gamma}}{4}-\frac{\pi_{\gamma}}{6}+\frac{L}{4}-\Phi\right)+k\Psi\nonumber\\
&&+\frac{1}{1+R}\left[{\cal H}\left(3c_s^2-1\right)V_{\rm b}+kc_s^2\left(\Delta_{\rm b}-3\Phi\right)-\dot{\cal V}\right]
\label{vgtc}
\end{eqnarray}
and
\begin{eqnarray}
\dot{V}_{\rm b}&=&\frac{1}{1+R}\left[{\cal H}\left(3c_s^2-1\right)V_{\rm b}+kc_s^2\left(\Delta_{\rm b}-3\Phi\right)\right]+k\Psi\nonumber\\
&& +\frac{R}{1+R}\left[k\left(\frac{\Delta_{\gamma}}{4}-\frac{\pi_{\gamma}}{6}+\frac{L}{4}-\Phi\right)+\dot{\cal V}\right].
\end{eqnarray}

\subsection{Initial conditions}
In the numerical calculation of the anisotropies in the CMB  it is usual to set the initial conditions after neutrino decoupling so that the neutrino anisotropic stress is non zero. 
This led to the formulation of initial conditions for the calculation of the CMB anisotropies in the presence of a primordial magnetic field in which the contributions involving the neutrino anisotropic stress and the magnetic stress cancel each other at lowest order in $x\equiv k\tau$ \cite{MG1, GK, fin1,fin2,le}. These are the compensating initial conditions. 
However, in a large class of models the magnetic field is generated in the very early universe, long before neutrino decoupling, such as during inflation (e.g. \cite{tw}) or the electroweak phase transition (e.g. \cite{ewp}).
In this case the magnetic field provides initially the only source of anisotropic stress.
It was shown in \cite{kkm} for a general type of anisotropic stress and in \cite{le,BC} for the special case of a magnetic field that the solution for the neutrino anisotropic stress after neutrino decoupling approaches a solution compensating the contribution from the anisotropic stress of the magnetic field.  Moreover,  there is an additional contribution to the curvature perturbation on large scales. 
Therefore in \cite{le} it was concluded that after neutrino decoupling there are effectively two types of perturbations associated with the primordial magnetic field. On the one hand there is  the compensating mode and on the other hand there is the passive mode which is an adiabatic-like mode with a non-vanishing curvature amplitude. 
Therefore the approach that is taken here to find the numerical solutions is to set 
the initial conditions after neutrino decoupling and assume  the compensating initial conditions. 

Following \cite{dmsw}  new variables are defined by, $\tilde{V}_i\equiv V_i/x$ and 
$\tilde{\pi}_i\equiv\pi_i/x^2$. Then the initial conditions on superhorizon scales, $x\ll 1$, are determined by the set of first order differential equations,
\begin{eqnarray}
\frac{d\Delta_{\gamma}}{d\ln x}&=&-\frac{4}{3}x^2\tilde{V}_{\gamma},\hspace{0.8cm}
\frac{d\Delta_{\nu}}{d\ln x}=-\frac{4}{3}x^2\tilde{V}_{\nu},\hspace{0.8cm}
\frac{d\Delta_{\rm c}}{d\ln x}=-x^2\tilde{V}_{\rm c},\hspace{0.8cm}
\frac{d\Delta_{\rm b}}{d\ln x}=-x^2\tilde{V}_{\rm b}\\
\frac{d\tilde{V}_{\gamma}}{d\ln x}&=&\frac{\Delta_{\gamma}}{4}-\tilde{V}_{\gamma}
+\Omega_{\gamma}\tilde{\pi}_{\rm B}+\Omega_{\nu}\tilde{\pi}_{\nu}
+2\Psi+\frac{L}{4},
\hspace{3.1cm}
\frac{d\tilde{V}_{\rm c}}{d\ln x}=-2\tilde{V}_{\rm c}+\Psi
\\
\frac{d\tilde{V}_{\nu}}{d\ln x}&=&\frac{\Delta_{\nu}}{4}-\tilde{V}_{\nu}+2\Psi+\Omega_{\gamma}\tilde{\pi}_{\rm B}+\Omega_{\nu}\tilde{\pi}_{\nu}-\frac{x^2}{6}\tilde{\pi}_{\nu}\\
\frac{d\tilde{\pi}_{\nu}}{d\ln x}&=&\frac{8}{5}\tilde{V}_{\nu}-2\tilde{\pi}_{\nu},
\hspace{7.2cm}
\frac{d\tilde{\pi}_{\rm B}}{d\ln x}=-2\tilde{\pi}_{\rm B}.
\end{eqnarray}
Solving this system to lowest order in $x$ imposing $\Delta_{\gamma}=\Delta_{\nu}=\frac{4}{3}\Delta_{\rm c}=\frac{4}{3}\Delta_{\rm b}$, the following set of initial conditions is found,
\begin{eqnarray}
\tilde{V}_{\nu}&=&-\frac{5}{4}\frac{\Delta_{\gamma}}{15+4\Omega_{\nu}}-\frac{5}{2}\frac{\Omega_{\gamma}\left(\Delta_{\rm B}+L\right)}{15+4\Omega_{\nu}}+\frac{5}{6}\frac{\Omega_{\gamma}}{\Omega_{\nu}}\frac{3-2\Omega_{\nu}}{15+4\Omega_{\nu}}\pi_{\rm B}\nonumber\\
\tilde{V}_{\gamma}&=&\tilde{V}_{\rm b}=-\frac{5}{4}\frac{\Delta_{\gamma}}{15+4\Omega_{\nu}}-\frac{5}{2}\frac{\Omega_{\gamma}\Delta_{\rm B}}{15+4\Omega_{\nu}}+\frac{5+14\Omega_{\nu}}{15+4\Omega_{\nu}}\frac{L}{4}-\frac{7}{3}\frac{\Omega_{\gamma}\pi_{\rm B}}{15+4\Omega_{\nu}}\nonumber\\
\tilde{V}_{\rm c}&=&-\frac{5}{4}\frac{\Delta_{\gamma}}{15+4\Omega_{\nu}}-\frac{5-4\Omega_{\nu}}{15+4\Omega_{\nu}}\frac{\Omega_{\gamma}}{8}\left(\Delta_{\rm B}+L\right)-\frac{13-4\Omega_{\nu}}{15+4\Omega_{\nu}}\frac{\Omega_{\gamma}\pi_{\rm B}}{12}\nonumber\\
\tilde{\pi}_{\nu}&=&-\frac{\Omega_{\gamma}}{\Omega_{\nu}}\tilde{\pi}_{\rm B}-\frac{\Delta_{\gamma}}{15+4\Omega_{\nu}}-\frac{2\Omega_{\gamma}\left(\Delta_{\rm B}+L\right)}{15+4\Omega_{\nu}}
+\frac{2}{3}\frac{\Omega_{\gamma}}{\Omega_{\nu}}\frac{3-2\Omega_{\nu}}{15+4\Omega_{\nu}}\pi_{\rm B}.
\label{ic}
\end{eqnarray}
As shown in the appendix, these  initial conditions correspond to the compensating initial conditions in the synchronous gauge used in previous numerical solutions \cite{MG1, GK, fin1,fin2,le}.
Moreover, in the case of no magnetic field these ("magnetized adiabatic") initial conditions reduce to the standard adiabatic initial conditions \cite{dmsw}.

The scalar curvature on a comoving hypersurface is given by (cf., e.g., \cite{RD})
\begin{eqnarray}
-\zeta=-\Phi+{\cal H}k^{-1}V.
 \label{curv}
 \end{eqnarray}
 The initial conditions are set after neutrino decoupling and we focus on the compensating magnetic mode. 
 Therefore, using equation (\ref{curv}), initially the total comoving curvature perturbation is given by, 
 \begin{eqnarray}
 \zeta=\frac{\Delta_{\gamma}}{4}+\Omega_{\gamma}\frac{\Delta_{\rm B}}{4}
 \label{zetand}
 \end{eqnarray}
  which can be used to express $\Delta_{\gamma}$ as $\Delta_{\gamma}=4\zeta-\Omega_{\gamma}\Delta_{\rm B}$. Moreover  $\zeta$ is treated as a Gaussian random variable with the two-point function in Fourier space $\langle\zeta^*(\vec{k})\zeta(\vec{k}')\rangle={\cal P}_{\zeta}\delta_{\vec«{k}\vec{k}'}$ and the dimensionless power spectrum is defined by ${\cal P}_{\zeta}=\frac{2\pi^2}{k^3}A_{\rm s}\left(\frac{k}{k_p}\right)^{n_s-1}$ where $n_s$ is the scalar spectral index, $A_s$ the amplitude of the scalar perturbations and $k_p=0.002$ Mpc$^{-1}$ the pivot wave number used in WMAP \cite{wmap7}.
In the numerical solutions $A_s$ and $n_s$ will be set to the bestfit values of the six parameter 
$\Lambda$CDM model of WMAP7 \cite{wmap7}. Taking into account that the initial total curvature perturbation is given by the sum of the contribution resulting, for example, from inflation and the magnetic contribution, assuming $\zeta$ to be determined by the bestfit values of WMAP and given the magnetic field parameters constrains the curvature perturbation from inflation.
 
 In \cite{le,BC} it was found that if the magnetic field is generated before neutrino decoupling the scalar curvature evolves on superhorizon scales to a final value at the time of neutrino 
 decoupling determined by the magnetic anisotropic stress, the time of generation of the magnetic field $\tau_{\rm B}$ and the time of neutrino decoupling $\tau_{\nu}$. This leads to the passive magnetic mode which corresponds to an adiabatic-like  mode with an amplitude  given by \cite{le},
 \begin{eqnarray}
 \zeta_{pass}\simeq-\frac{1}{3}R_{\gamma}\pi_{\rm B}\left[\log\left(\frac{\tau_{\nu}}{\tau_{\rm B}}\right)+\frac{5}{8R_{\nu}}-1\right],
 \label{zeta1}
 \end{eqnarray}
 where $R_{\gamma}\equiv\frac{\Omega_{\gamma}}{\Omega_{\gamma}+\Omega_{\nu}}$. 
 If the magnetic field is generated during a phase transition, which is the case considered by \cite{BC} then $\tau_{\rm B}$ corresponds to the time of the phase transition, if however, it is generated during inflation  $\tau_{\rm B}$ could be chosen to be at reheating \cite{le}.
 The passive mode has to be added to equation (\ref{zetand}) which consequently modifies the constraint on the nonmagnetic contribution to the initial total curvature perturbation.

\section{The magnetic field contribution}
\label{sec3}
\setcounter{equation}{0}
In order to describe the magnetic field the so called "lab" frame is chosen, in which 
\begin{eqnarray}
B_i(\vec{x},\tau)=\frac{1}{2a^2}\sum_{j,m}\epsilon_{ijm}F_{jm}, 
\end{eqnarray}
where $\epsilon_{ijm}$ is the totally 
antisymmetric symbol, with $\epsilon_{123}=1$ and $F_{\mu\nu}$ is the Maxwell tensor.
The lab frame is defined by choosing locally Minkowski space-time, so that the lab  coordinates are defined by $dt=ad\tau$, $d\vec{r}=ad\vec{x}$ \cite{sb}.
Assuming that the conductivity of the plasma is large, Ohm's law implies a vanishing electric field, which leads to the magnetic field decaying as $1/a^2$ as the universe expands.
In general the energy momentum tensor of the electromagnetic field measured by the fundamental observer can be written in terms of that of an imperfect fluid \cite{tsagas},
\begin{eqnarray}
T_{\alpha\beta}=(\rho+p)u_{\alpha}u_{\beta}+pg_{\alpha\beta}+2u_{(\alpha}q_{\beta)}+\pi_{\alpha\beta},
\end{eqnarray}
where $u^{\alpha}=a^{-1}\delta^{\alpha}_0$ is the 4-velocity of the fluid and $u_{\alpha}u^{\alpha}=-1$. $q_{\alpha}$ is the heat flux which, in the case of the electromagnetic field is determined by the Poynting vector and thus vanishes for vanishing electric field \cite{tsagas}.
The  magnetic energy density $\rho_{\rm B}$, pressure $p_{\rm B}$ and anisotropic stress 
$\pi_{\rm (B)\;\alpha\beta}$ in the lab frame are given by \cite{tsagas},
\begin{eqnarray}
\rho_{\rm B}=\frac{\vec{B}^2(\vec{x},\tau)}{2},\hspace{1.8cm}p_{\rm B}=\frac{1}{3}\rho_{\rm B},
\hspace{1.8cm}
\pi_{{\rm (B)}\;ij}=-B_i(\vec{x},\tau)B_j(\vec{x},\tau)+\frac{1}{3}\vec{B}^2(\vec{x},\tau)\delta_{ij},
\label{p1}
\end{eqnarray}
where the anisotropic stress has only non-vanishing spatial components and the vector notation denotes a spatial 3-vector. Moreover, it has been used that the anisotropic stress tensor changes from the FRW frame to the lab frame as $\pi_{{\rm B}\;ij}^{\rm FRW}=\pi_{{\rm B}\;ij}^{\rm lab}a^2$.
Furthermore, the term due to the Lorentz force entering the equation of the baryon velocity evolution and in the tight-coupling limit the photon velocity evolution which is derived from $\nabla^{\alpha}T^{\rm (em)}_{\alpha\beta}=-F_{\beta\alpha}J^{\alpha}$ and expressed in terms of quantities in the lab frame yields to
\begin{eqnarray}
\vec{L}(\vec{x},\tau)=a\left(\vec{J}\times\vec{B}\right)(\vec{x},\tau)
\end{eqnarray}
For vanishing electric field the current $\vec{J}$ is given by
$a\vec{J}=\nabla\times\vec{B}$. Thus the components of the  Lorentz term  takes the form
\begin{eqnarray}
L_j=-\frac{1}{6}\partial_j\vec{B}^2-\sum_{i}\partial_i\pi_{{\rm (B)}\;ij}
\end{eqnarray}

In the perturbation equations the magnetic field contributes to the total energy contrast and the total anisotropic stress.
The magnetic energy density $\rho_{\rm B}$ defines the magnetic energy density contrast $\delta_{\rm B}$ as
\begin{eqnarray}
\rho_{\rm B}(\vec{x},\tau)=\rho_{\gamma}\sum_{\vec{k}}\delta_{\rm B}(\vec{k}) Y(\vec{k},\vec{x}),
\end{eqnarray}
 where $Y(\vec{k},\vec{x})$ denote a complete set of scalar harmonic functions satisfying, $(\triangle+k^2)Y=0$ \cite{ks}. The magnetic energy density is defined in such a way that it does not contribute to the total  background  energy density. This is different from \cite{tsagas} where the magnetic energy density does contribute to the background. Moreover, $\Delta_{\rm B}=\delta_{\rm B}$.
The anisotropic stress can be expanded in terms of scalar harmonics as,
\begin{eqnarray}
\pi_{{\rm (B)}\;ij}=p_{\gamma}\sum_{\vec{k}}\pi_{\rm B}(\vec{k})Y_{ij}(\vec{k},\vec{x}),
\label{p2}
\end{eqnarray}
where $Y_{ij}=k^{-2}Y_{|ij}+\frac{1}{3}\delta_{ij}Y$ \cite{ks}.
Thus the Lorentz term can be written as
\begin{eqnarray}
L_i(\vec{x},\tau)=\frac{\rho_{\gamma}}{3}\sum_{\vec{k}}kL(\vec{k})Y_i(\vec{k},\vec{x})
\hspace{2cm}
L(\vec{k})=\Delta_{\rm B}-\frac{2}{3}\pi_{\rm B}
\label{lorentz}
\end{eqnarray}
and $Y_i\equiv -k^{-1}Y_{|i}$ \cite{ks}.
The magnetic energy density, the anisotropic stress and the Lorentz term in Fourier space can be related to the magnetic field spectrum.
Using that $B_i(\vec{x},\tau)=B_i(\vec{x},\tau_0)\left(\frac{a_0}{a(\tau)}\right)^2$ and 
$\rho_{\gamma}=\rho_{\gamma\,0}\left(\frac{a_0}{a}\right)^4$, where the index 0 refers to the present epoch and defining
\begin{eqnarray}
B_i(\vec{x},\tau_0)=\sum_{\vec{k}}B_i(\vec{k})Y(\vec{k},\vec{x}),
\label{p3}
\end{eqnarray}
where $B_i(\vec{k})\equiv B_i(\vec{k},\tau_0)$.
Thus it is found that 
\begin{eqnarray}
\Delta_{\rm B}(\vec{k})=\frac{1}{2\rho_{\gamma\,0}}\sum_{\vec{q}}B_i(\vec{q})B^i(\vec{k}-\vec{q})
\end{eqnarray}
and using the last equation of  (\ref{p1}) and equations (\ref{p2}) and (\ref{p3}) yields
\begin{eqnarray}
\pi_{\rm B}(\vec{k})=\frac{3}{2\rho_{\gamma\,0}}\left[\sum_{\vec{q}}\frac{3}{k^2}B_i(\vec{q})(k^i-q^i)
B_j(\vec{k}-\vec{q})q^j-\sum_{\vec{q}}B_m(\vec{q})B^m(\vec{k}-\vec{q})\right].
\end{eqnarray}
The magnetic field at present  is characterized by its spectrum $P_{\rm B}$ which is chosen to be of the form
\begin{eqnarray}
P_{\rm B}(k,k_{\rm m},k_{\rm L})=A_{\rm B}\left(\frac{k}{k_{\rm L}}\right)^{n_{\rm B}}W(k,k_{\rm m}),
\label{P}
\end{eqnarray}
where $A_{B}$ is its amplitude, $n_{\rm B}$ is the magnetic spectral index and $k_{\rm m}$ the upper cut-off in the magnetic field spectrum due to diffusion of the magnetic field energy density on small scales. $k_{\rm L}$ is the pivot scale of the magnetic field and $W(k,k_{\rm m})$ is the window function modeling the cut-off of the magnetic field spectrum due to diffusion.
As shown in \cite{sb} the damping of the magnetic field is determined by the dimensionless Alfv\'en velocity $V_{\rm A}$ and the Silk damping scale $k_{\rm S}$, that is
\begin{eqnarray}
k_{\rm m}^{-2}=V_{\rm A}^2k_{\rm S}^{-2},
\end{eqnarray}
where $V_{\rm A}\simeq 3.8\times 10^{-4}(B/1\, {\rm nG})$ and  $B$ corresponds to the  smoothed field strength of the  magnetic field today and 
before recombination the Silk scale may be approximated by \cite{hs1}
\begin{eqnarray}
k_{\rm S}^{-2}\simeq 1.7\times 10^7\left(1-\frac{Y_{\rm P}}{2}\right)^{-1}(\Omega_{\rm b}h^2)^{-1}
(\Omega_0 h^2)^{-\frac{1}{2}}\left(\frac{a}{a_0}\right)^{\frac{5}{2}}\frac{1}{3\sqrt{a_{\rm eq}/a}+2}
{\rm Mpc}^2,
\end{eqnarray}
where $Y_{\rm P}$ is the primordial helium mass fraction.
After decoupling the photon mean free path becomes infinite and the viscous damping becomes subdominant. Therefore, the largest scale of the magnetic field spectrum damped corresponds to $k_m^{-1}$ evaluated at recombination. Thus, it is interesting to note that
the diffusion scale changes with time and this would lead to a magnetic field spectrum which has a more general time dependence than just the decay due to expansion.  Therefore, since this case is not considered here, the magnetic damping scale is assumed to be defined by the largest damped scale.
This scale is determined by the Alfv\'en velocity and the Silk damping scale at recombination.
The Silk damping scale is determined by the mean free path of the photons which continuously grows from the early epochs within the tight-coupling regime of the baryon-photon fluid to recombination shortly after which the photon mean free path becomes infinity.
In \cite{sb} the maximal damped magnetic scale has been estimated and the corresponding maximal wave number is given by 
\begin{eqnarray}
k_{\rm m}\simeq 200.694\left(\frac{B}{\rm nG}\right)^{-1}{\rm Mpc}^{-1},
\label{km}
\end{eqnarray}
 for the values of the bestfit $\Lambda$CDM model of  WMAP7, $\Omega_b=0.0227h^{-2}$ and $h=0.714$ \cite{wmap7}.
 
\begin{figure}[h!]
\centerline{\epsfxsize=2.in\epsfbox{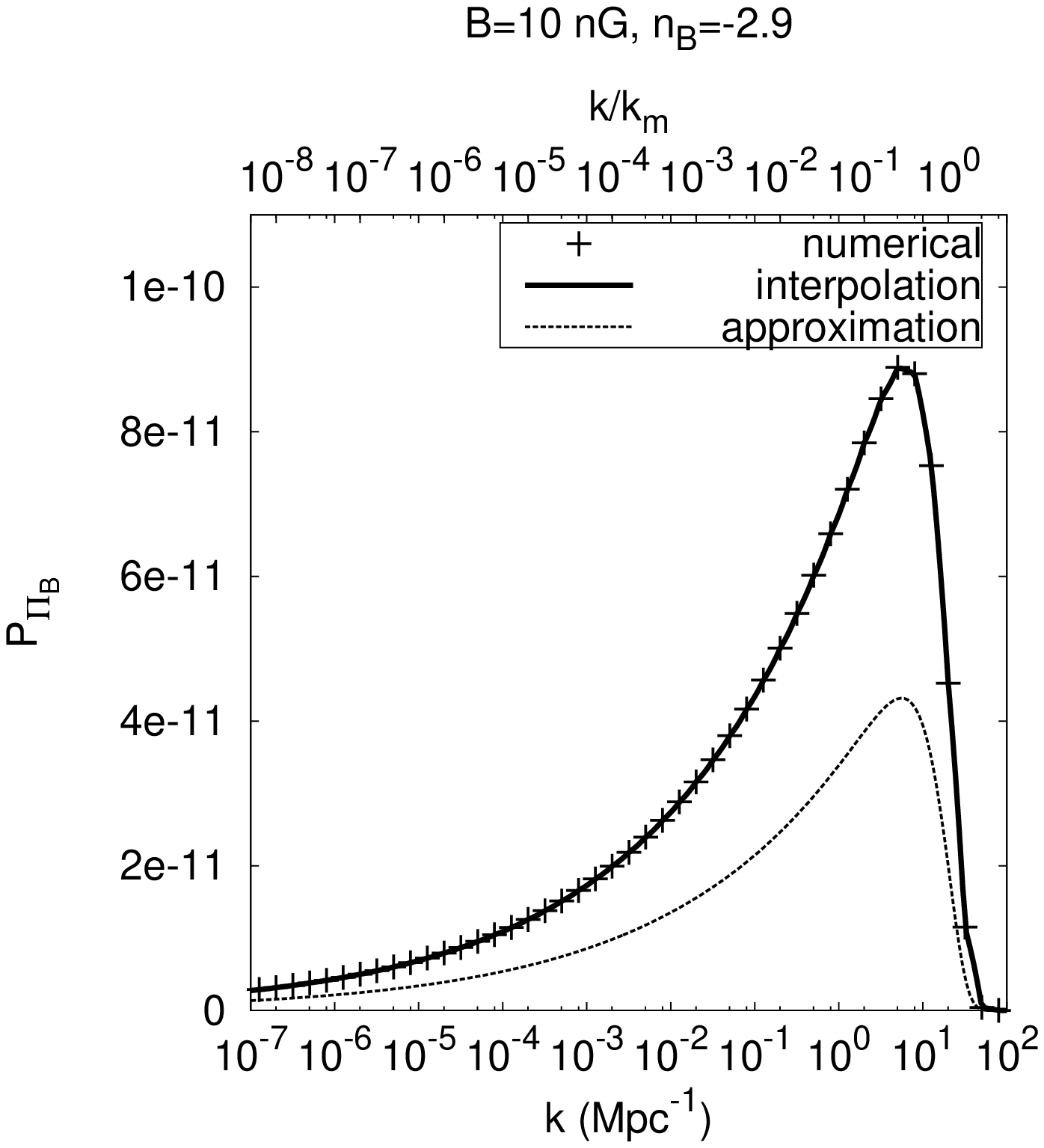}
\hspace{0.5cm}
\epsfxsize=2.in\epsfbox{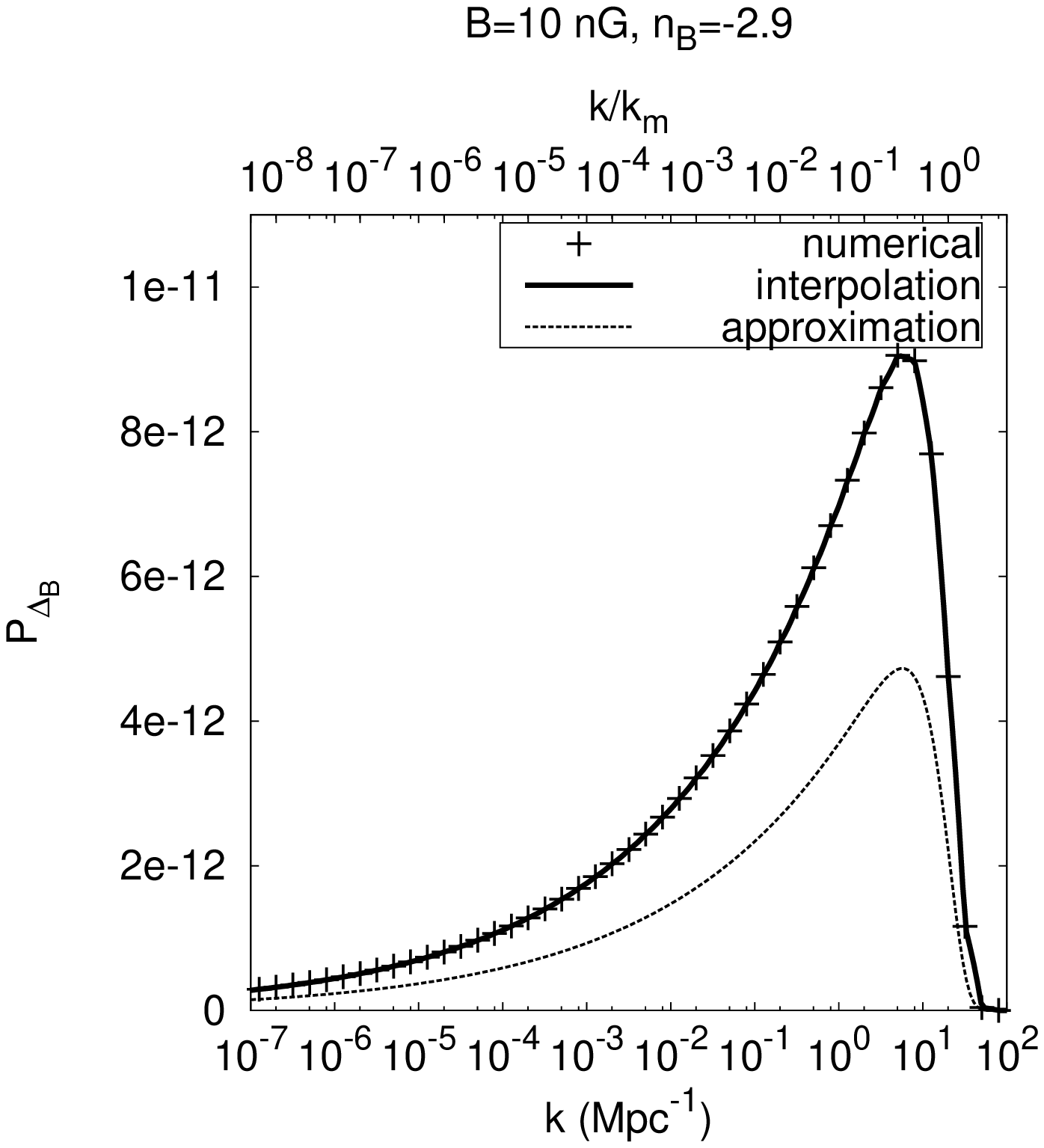}\hspace{0.5cm}
\epsfxsize=2.in\epsfbox{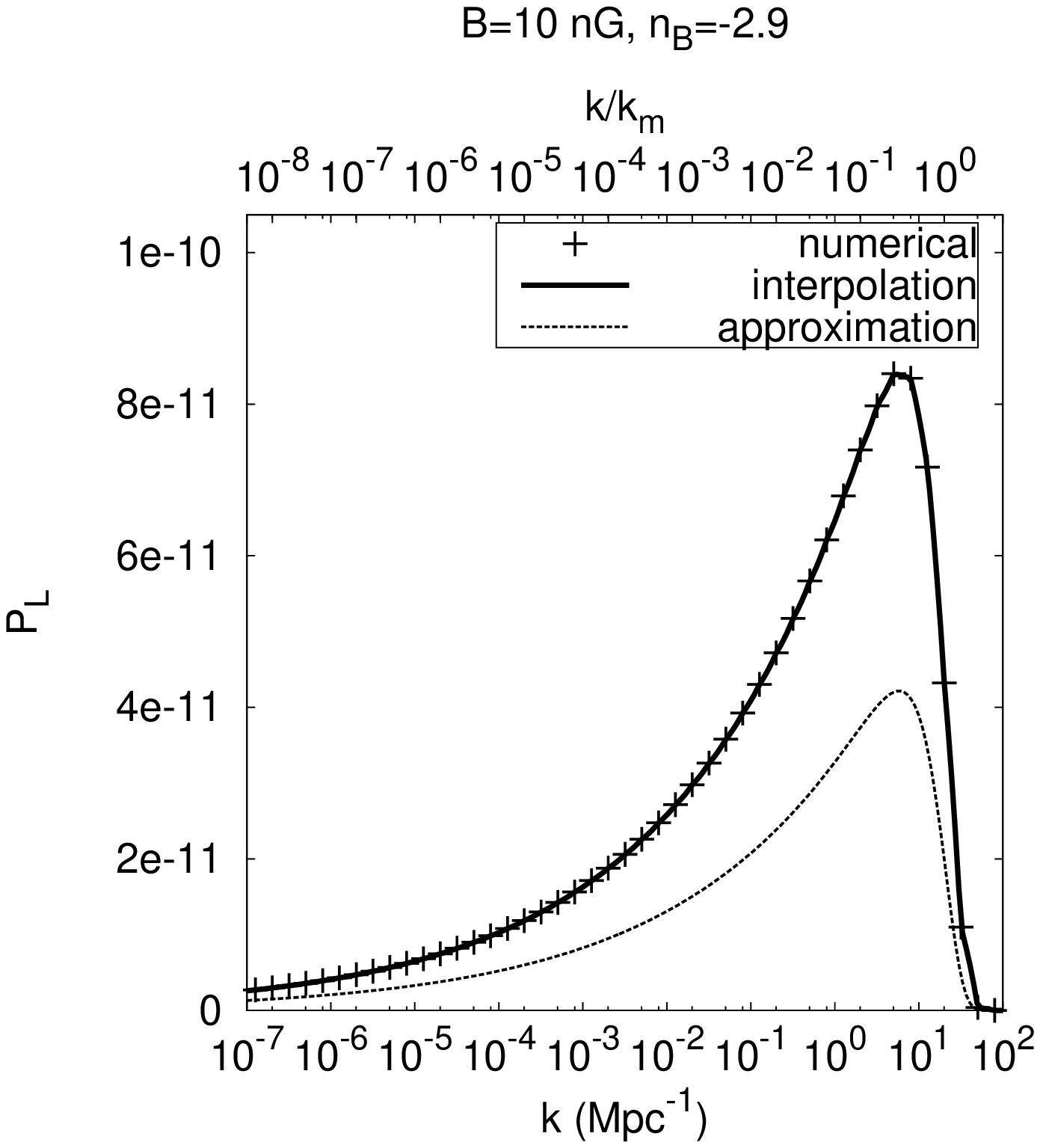 }}
\caption{The spectra determining the autocorrelation functions of the magnetic anisotropic stress, energy density contrast and the Lorentz term are shown. For $B=10$ nG and spectral index $n_{\rm B}=-2.9$ in each graph the numerical solution of the double integral, together with the numerical interpolation (spline) used in the code to calculate the CMB anisotropies and the analytical approximation in terms of incomplete Gamma functions is shown. The lower horizontal axis shows $k$, the upper one shows the ratio $k/k_{\rm m}$. In this example, the magnetic damping wave number is given by  $k_{\rm m}=20$ Mpc$^{-1}$ (cf. equation (\ref{km})). }
\label{fig1}
\end{figure}

 In the following the window function is assumed to be gaussian of the form,
\begin{eqnarray}
W(k,k_{\rm m})=\pi^{-\frac{3}{2}}k_{m}^{-3}e^{-\left(\frac{k}{k_{\rm m}}\right)^2},
\label{w}
\end{eqnarray}
in such a way that $\int d^3k W(k,k_{\rm m})=1$. This choice of window function is different from the step function used in previous work \cite{GK, fin1,fin2,le}.
The smoothed magnetic field strength is defined by the auto-correlation function,
$\langle \vec{B}^2(\vec{x},\tau_0)\rangle$
which can be easily calculated using  the correlation function of the magnetic field given by
\begin{eqnarray}
\langle B_i^*(\vec{k})B_j(\vec{q})\rangle=\delta_{\vec{k},\vec{q}}\,P_{\rm B}\left(\delta_{ij}-\frac{k_ik_j}{k^2}\right),
\end{eqnarray}
where the helical part is neglected since for a stochastic magnetic field it does not contribute to the CMB anisotropies due to scalar perturbations \cite{heli}.
Taking the continuum limit, $\sum_{\vec{k}}\rightarrow \int\frac{d^3k}{(2\pi)^3}$ implies that the 
magnetic field strength today smoothed over the magnetic diffusion scale is given by 
\begin{eqnarray}
\langle \vec{B}(\vec{x})^2\rangle=A_{\rm B}\pi^{-\frac{7}{2}}\left(\frac{k_{\rm m}}{k_{\rm L}}\right)^{n_{\rm B}}\frac{\Gamma\left(\frac{n_{\rm B}+3}{2}\right)}{2},
\end{eqnarray}
which is valid for $n_{\rm B}>-3$. The damping scale is a natural scale in the problem. Therefore here it is not necessary to introduce a smoothing scale as it is usually done. 
Using equation (\ref{km}) the damping scale $\lambda_{\rm m}=2\pi/k_{\rm m}$ is of the order of $\lambda_{\rm m}\simeq 30 (B/{\rm nG})$ kpc and thus in general smaller than 1 Mpc which is usually used as smoothing scale. 
Defining the two-point correlation functions in $k$-space in terms of the dimensionless spectrum ${\cal P}_F$ by 
\begin{eqnarray}
\langle F_{\vec{k}}^*F_{\vec{k}'}\rangle=\frac{2\pi^2}{k^3}{\cal P}_{F}(k)\delta_{\vec{k},\vec{k}'}
\label{F}
\end{eqnarray}
The following expressions are found for the spectra determining the auto correlation functions of the magnetic energy density contrast, the anisotropic stress and the Lorentz term. These are different from previously obtained expressions \cite{fin1,fin2, le,sl10} due to the different choice of window function.
In general the spectrum characterizing the auto-correlation function of the magnetic energy density contrast is given by \cite{magstoc},
 \begin{eqnarray}
 {\cal P}_{\Delta_{\rm B}}(k)=\frac{k^3}{4\pi^2\rho_{\gamma\, 0}^2}\sum_{\vec{q}}P_{\rm B}(q)P_{\rm B}(|\vec{k}-\vec{q}|)\left[1+\frac{\left[\vec{q}\cdot(\vec{k}-\vec{q})\right]^2}{q^2|\vec{k}-\vec{q}|^2}\right]
 \end{eqnarray}
 which for the spectral function (\ref{P}) leads to 
\begin{eqnarray}
{\cal P}_{\Delta_{\rm B}}(k,k_{\rm m})&=&\frac{1}{\left[\Gamma\left(\frac{n_{\rm B}+3}{2}\right)\right]^2}
\left[\frac{\rho_{\rm B\,0}}{\rho_{\gamma\, 0}}\right]^2\left(\frac{k}{k_{\rm m}}\right)^{2(n_{\rm B}+3)}e^{-\left(\frac{k}{k_{\rm m}}\right)^2}\int_0^{\infty}dz z^{n_{\rm B}+2}e^{-2\left(\frac{k}{k_{\rm m}}\right)^2z^2}\nonumber\\
&&\hspace{1cm}
\int_{-1}^1 dxe^{2\left(\frac{k}{k_{\rm m}}\right)^2zx}\left(1-2zx+z^2\right)^{\frac{n_{\rm B}-2}{2}}\left(1+x^2+2z^2-4zx\right),
\label{PD}
\end{eqnarray}
where $x\equiv \vec{k}\cdot\vec{q}/(kq)$ and $z\equiv\frac{q}{k}$.
Moreover, the average energy density of the magnetic field is defined to be $\rho_{{\rm B}\,0}=\langle\vec{B}^2(\vec{x})\rangle/2$.
For small values of $\frac{k}{k_{\rm m}}$, that is $\frac{k}{k_{\rm m}}\ll 1$,  the double integral can be approximated by incomplete Gamma functions.  In particular, it is found that
\begin{eqnarray}
{\cal P}_{\Delta_{\rm B}}(k,k_{\rm m})&=&\frac{1}{\left[\Gamma\left(\frac{n_{\rm B}+3}{2}\right)\right]^2}
\left[\frac{\rho_{\rm B\,0}}{\rho_{\gamma\, 0}}\right]^2\left(\frac{k}{k_{\rm m}}\right)^{2(n_{\rm B}+3)}e^{-\left(\frac{k}{k_{\rm m}}\right)^2}
\times\nonumber\\
&&\left[\frac{2^{\frac{1-n_{\rm B}}{2}}}{3}\left(\frac{k}{k_{\rm m}}\right)^{-(n_{\rm B}+3)}\gamma\left(\frac{n_{\rm B}+3}{2};2\left(\frac{k}{k_{\rm m}}\right)^2\right)
\right.
\nonumber\\
&+&
\left.2^{-(n_{\rm B}+\frac{1}{2})}\left(\frac{k}{k_{\rm m}}\right)^{-2n_{\rm B}-3}\Gamma\left(n_{\rm B}+\frac{3}{2};2\left(\frac{k}{k_{\rm m}}\right)^2\right)\right].
\label{appD}
\end{eqnarray}
where $\gamma(\alpha,x)\equiv\int_0^xe^{-t}t^{\alpha-1}dt$ and $\Gamma(\alpha,x)\equiv\int_x^{\infty}e^{-t}t^{\alpha-1}dt$ are incomplete Gamma functions \cite{grad}.
It is interesting to note that in the limit $k\ll k_{\rm m}$ the power-law behaviour which was used in
the magnetic field spectra used in \cite{GK} is recovered. However, as can be seen from figure 
\ref{fig1} the approximation underestimates the contribution from the magnetic field and thus the  effect on the final angular power spectrum of the temperature anisotropies and polarization.

\begin{figure}[h!]
\centerline{\epsfxsize=2.in\epsfbox{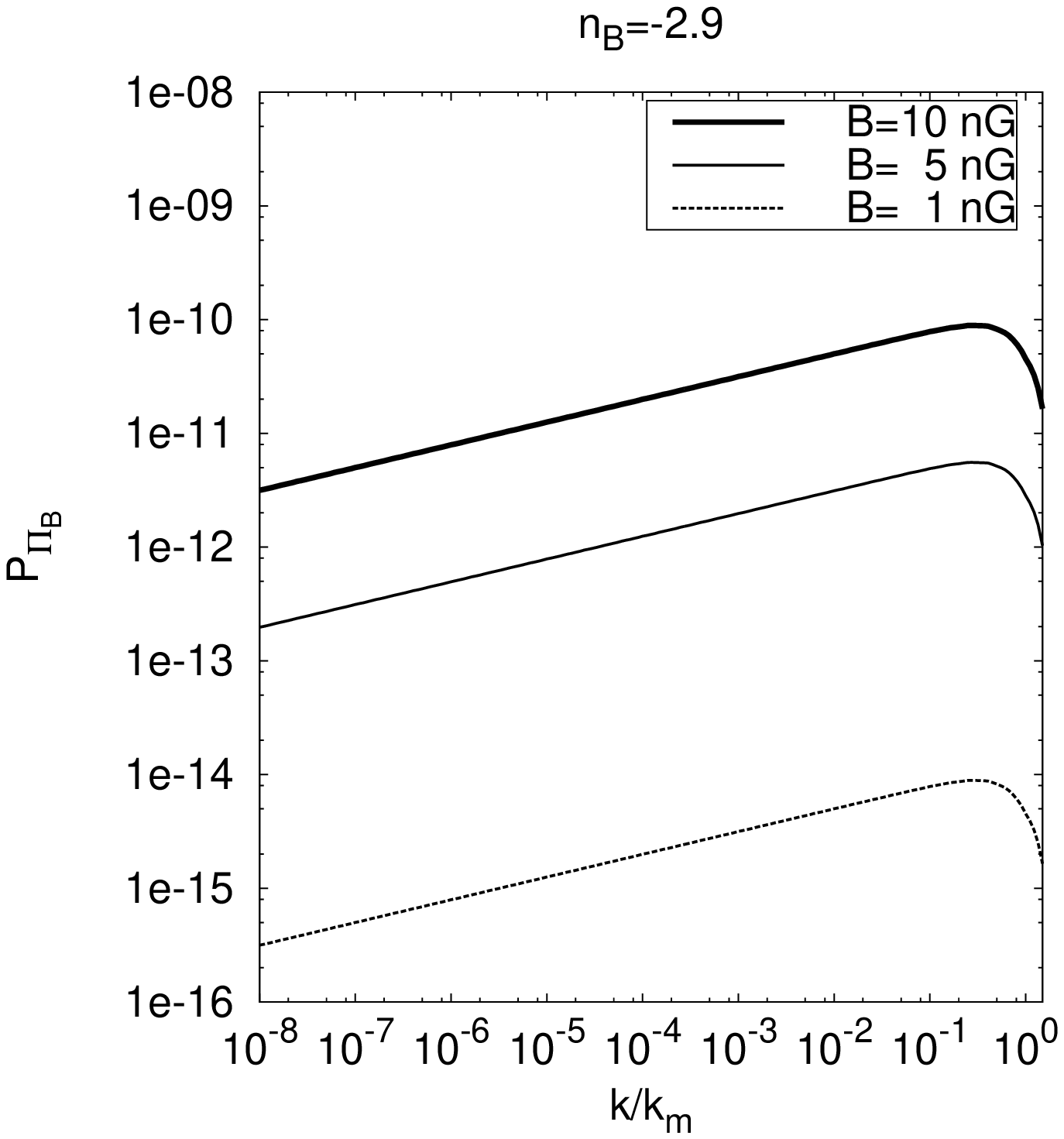}
\hspace{0.5cm}
\epsfxsize=2.in\epsfbox{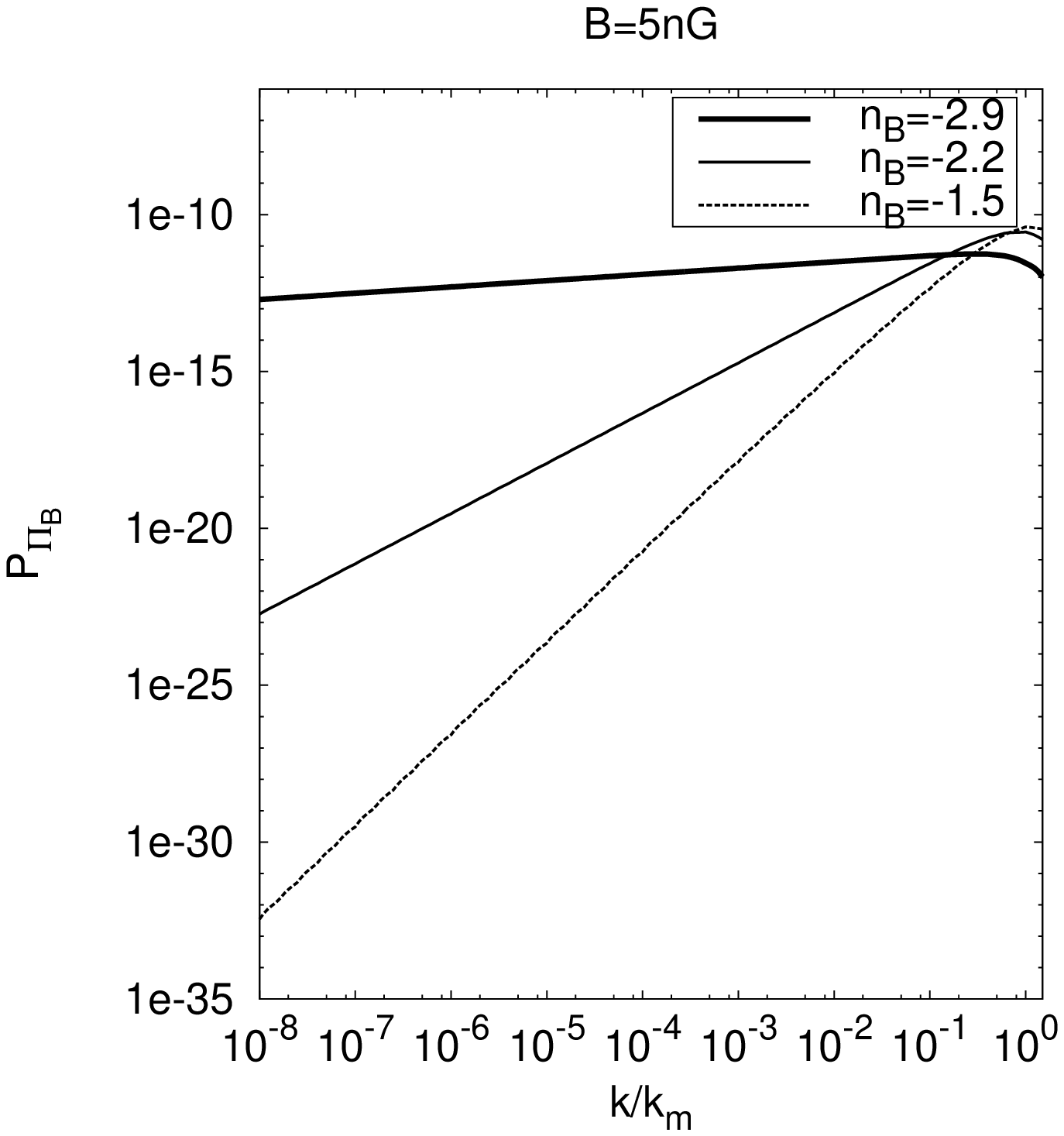}}
\caption{The spectra determining the autocorrelation function of the magnetic anisotropic stress ${\cal P}_{\pi_{\rm B}}$ are shown for different values of the magnetic field strength $B$ and magnetic spectral index $n_{\rm B}$. In this case the numerical interpolation of the double integral appearing in equation (\ref{Ppp}) was used. The magnetic damping wave number $k_{\rm m}$ is given by equation (\ref{km}).}
\label{fig2}
\end{figure}

The spectrum defining the auto correlation function of the anisotropic stress is determined by \cite{magstoc},
\begin{eqnarray}
{\cal P}_{\pi_{\rm B}}(k,k_{\rm m})&=&\frac{9k^3}{8\pi^2\rho_{\gamma\, 0}^2}\sum_{\vec{q}}
P_{\rm B}(q)P_{\rm B}(|\vec{k}-\vec{q}|)\left[\frac{18q^2(1-x^2)^2}{|\vec{k}-\vec{q}|^2}+12\frac{q(kx-q)(1-x^2)}{|\vec{k}-\vec{q}|^2}
\right.\nonumber\\
&&\left.\hspace{5cm}
+2\left(1+\frac{(kx-q)^2}{|\vec{k}-\vec{q}|^2}\right)\right].
\end{eqnarray}
Using the expression for the magnetic field spectrum (\ref{P}) and the window function (\ref{w})
this leads in the continuum limit to,
\begin{eqnarray}
{\cal P}_{\pi_{\rm B}}(k,k_{\rm m})=\frac{9}{\left[\Gamma\left(\frac{n_{\rm B}+3}{2}\right)\right]^2}
\left[\frac{\rho_{\rm B\,0}}{\rho_{\gamma\, 0}}\right]^2\left(\frac{k}{k_{\rm m}}\right)^{2(n_{\rm B}+3)}e^{-\left(\frac{k}{k_{\rm m}}\right)^2}\int_0^{\infty}dz z^{n_{\rm B}+2}e^{-2\left(\frac{k}{k_{\rm m}}\right)^2z^2}\nonumber\\
\int_{-1}^1 dxe^{2\left(\frac{k}{k_{\rm m}}\right)^2zx}\left(1-2zx+z^2\right)^{\frac{n_{\rm B}-2}{2}}
\left(1+5z^2+2zx+(1-12z^2)x^2
-6zx^3+9z^2x^4\right).
\label{Ppp}
\end{eqnarray}
This gives in the limit for $\frac{k}{k_{\rm m}}\ll 1$, 
\begin{eqnarray}
{\cal P}_{\pi_{\rm B}}(k,k_{\rm m})&=&\frac{9}{\left[\Gamma\left(\frac{n_{\rm B}+3}{2}\right)\right]^2}
\left[\frac{\rho_{\rm B\,0}}{\rho_{\gamma\, 0}}\right]^2\left(\frac{k}{k_{\rm m}}\right)^{2(n_{\rm B}+3)}e^{-\left(\frac{k}{k_{\rm m}}\right)^2}\times
\nonumber\\
&&\left[\frac{2^{\frac{1-n_{\rm B}}{2}}}{3}\left(\frac{k}{k_{\rm m}}\right)^{-(n_{\rm B}+3)}\gamma\left(\frac{n_{\rm B}+3}{2};2\left(\frac{k}{k_{\rm m}}\right)^2\right)
\right.
\nonumber\\
&+&
\left.
\frac{7}{5}2^{-(n_{\rm B}+\frac{1}{2})}\left(\frac{k}{k_{\rm m}}\right)^{-2n_{\rm B}-3}\Gamma\left(n_{\rm B}+\frac{3}{2};2\left(\frac{k}{k_{\rm m}}\right)^2\right)\right].
\label{appP}
\end{eqnarray}
Finally, the autocorrelation function for the Lorentz term is determined by,
\begin{eqnarray}
{\cal P}_{L}(k,k_{\rm m})&=&\frac{9k^3}{4\pi^2\rho_{\gamma\, 0}^2}\sum_{\vec{q}}
P_{\rm B}(q)\frac{P_{\rm B}(|\vec{k}-\vec{q}|)}{|\vec{k}-\vec{q}|^2}\left[k^2-2kqx+q^2+(kx-q)^2
\right.\nonumber\\
&&\left.\hspace{4cm}
+4q(kx-q)(1-x^2)+4q^2(1-x^2)^2\right],
\end{eqnarray}
which for the case at hand in the continuum limit leads to,
\begin{eqnarray}
{\cal P}_L(k,k_{\rm m})=\frac{9}{\left[\Gamma\left(\frac{n_{\rm B}+3}{2}\right)\right]^2}
\left[\frac{\rho_{\rm B\,0}}{\rho_{\gamma\, 0}}\right]^2\left(\frac{k}{k_{\rm m}}\right)^{2(n_{\rm B}+3)}e^{-\left(\frac{k}{k_{\rm m}}\right)^2}\int_0^{\infty}dz z^{n_{\rm B}+2}e^{-2\left(\frac{k}{k_{\rm m}}\right)^2z^2}\nonumber\\
\int_{-1}^1 dxe^{2\left(\frac{k}{k_{\rm m}}\right)^2zx}\left(1-2zx+z^2\right)^{\frac{n_{\rm B}-2}{2}}
\left[1+2z^2+(1-4z^2)x^2-4zx^3+4z^2x^4\right].
\end{eqnarray}
In the limit $\frac{k}{k_{\rm m}}\ll 1$ this can be approximated by,
\begin{eqnarray}
{\cal P}_{L}(k,k_{\rm m})&=&\frac{9}{\left[\Gamma\left(\frac{n_{\rm B}+3}{2}\right)\right]^2}
\left[\frac{\rho_{\rm B\,0}}{\rho_{\gamma\, 0}}\right]^2\left(\frac{k}{k_{\rm m}}\right)^{2(n_{\rm B}+3)}e^{-\left(\frac{k}{k_{\rm m}}\right)^2}\
\times\nonumber\\
&&
\left[\frac{2^{\frac{1-n_{\rm B}}{2}}}{3}\left(\frac{k}{k_{\rm m}}\right)^{-(n_{\rm B}+3)}\gamma\left(\frac{n_{\rm B}+3}{2};2\left(\frac{k}{k_{\rm m}}\right)^2\right)
\right.
\nonumber\\
&+&
\left.
\frac{11}{15}2^{-(n_{\rm B}+\frac{1}{2})}\left(\frac{k}{k_{\rm m}}\right)^{-2n_{\rm B}-3}\Gamma\left(n_{\rm B}+\frac{3}{2};2\left(\frac{k}{k_{\rm m}}\right)^2\right)\right].
\label{appL}
\end{eqnarray}
It is interesting to note that the spectral functions only depend on the ratio $k/k_{\rm m}$.
In figure \ref{fig1} the correlation functions are shown.
In figure \ref{fig2} the spectrum determining the autocorrelation function of the anisotropic stress is shown for various values of the magnetic field strength and its spectral index.
The approximate analytical solutions capture the general shape of the different power spectra, however, the amplitude is in all cases far below the one of the  numerical solution. Only for wave numbers $k$ much smaller than the one corresponding to the damping scale is the difference in the two curves not significant.

\section{Results}

\begin{figure}[h!]
\centerline{\epsfxsize=2.2in\epsfbox{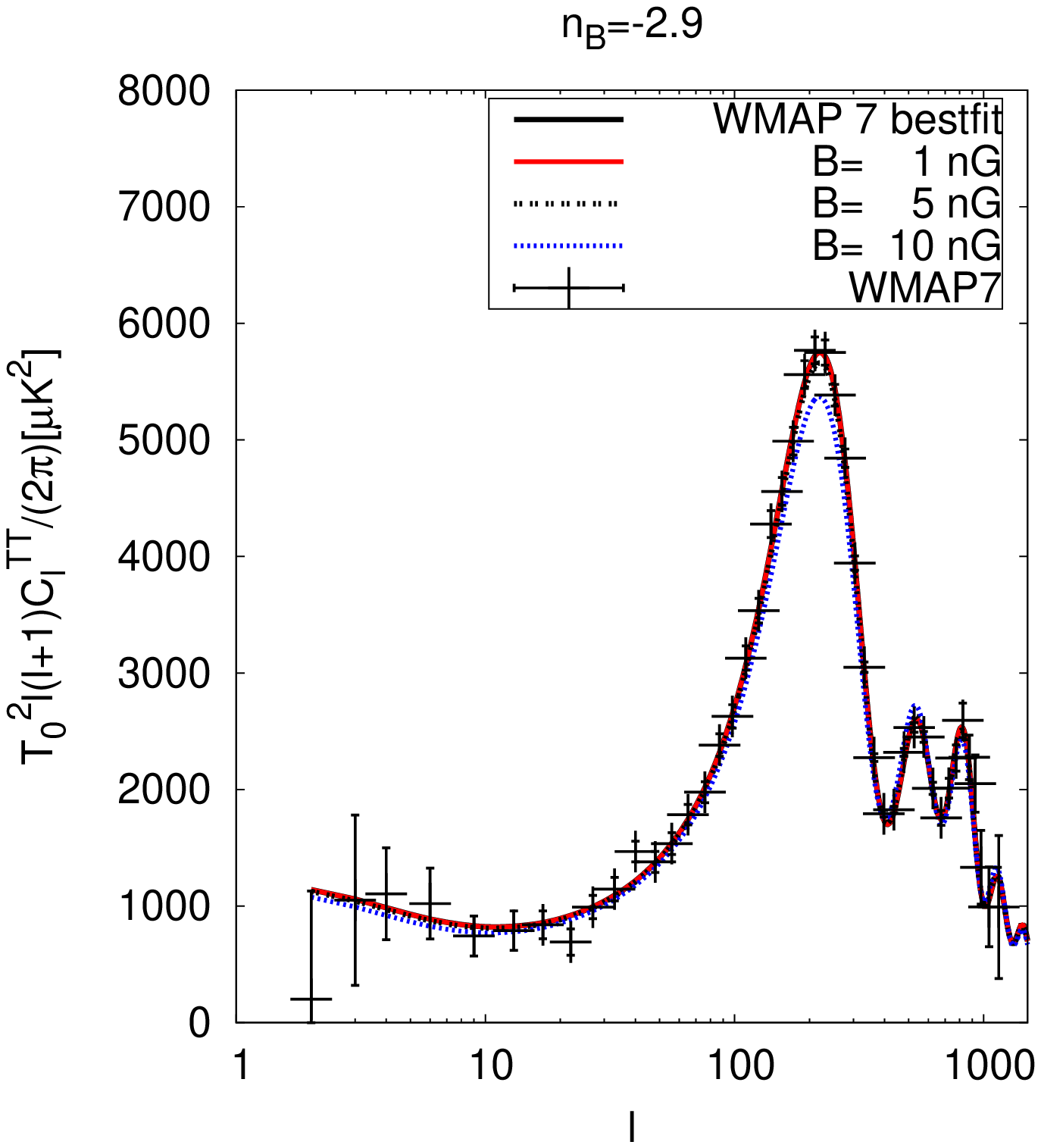}
\hspace{0.05cm}
\epsfxsize=2.2in\epsfbox{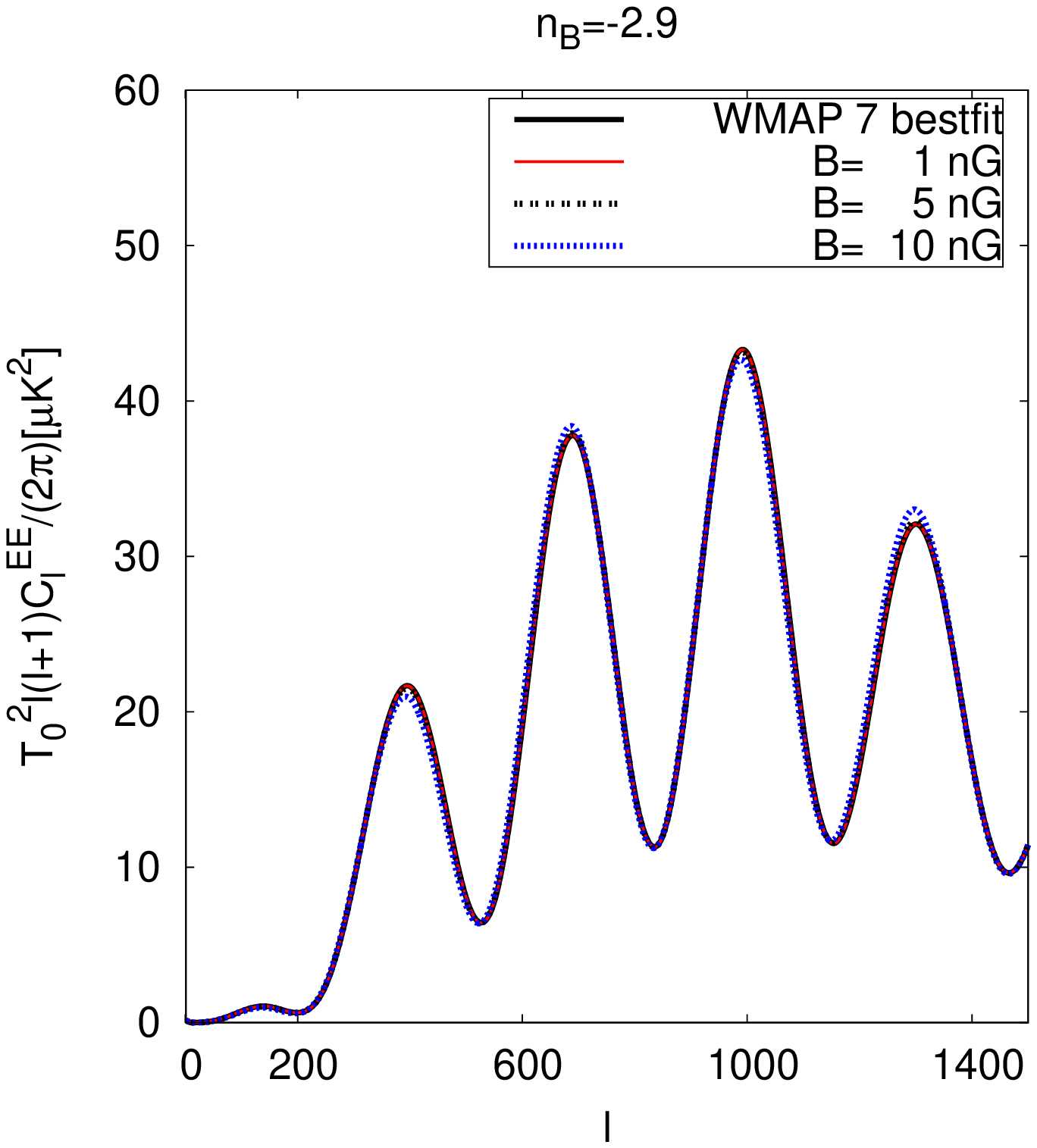}\hspace{0.05cm}
\epsfxsize=2.2in\epsfbox{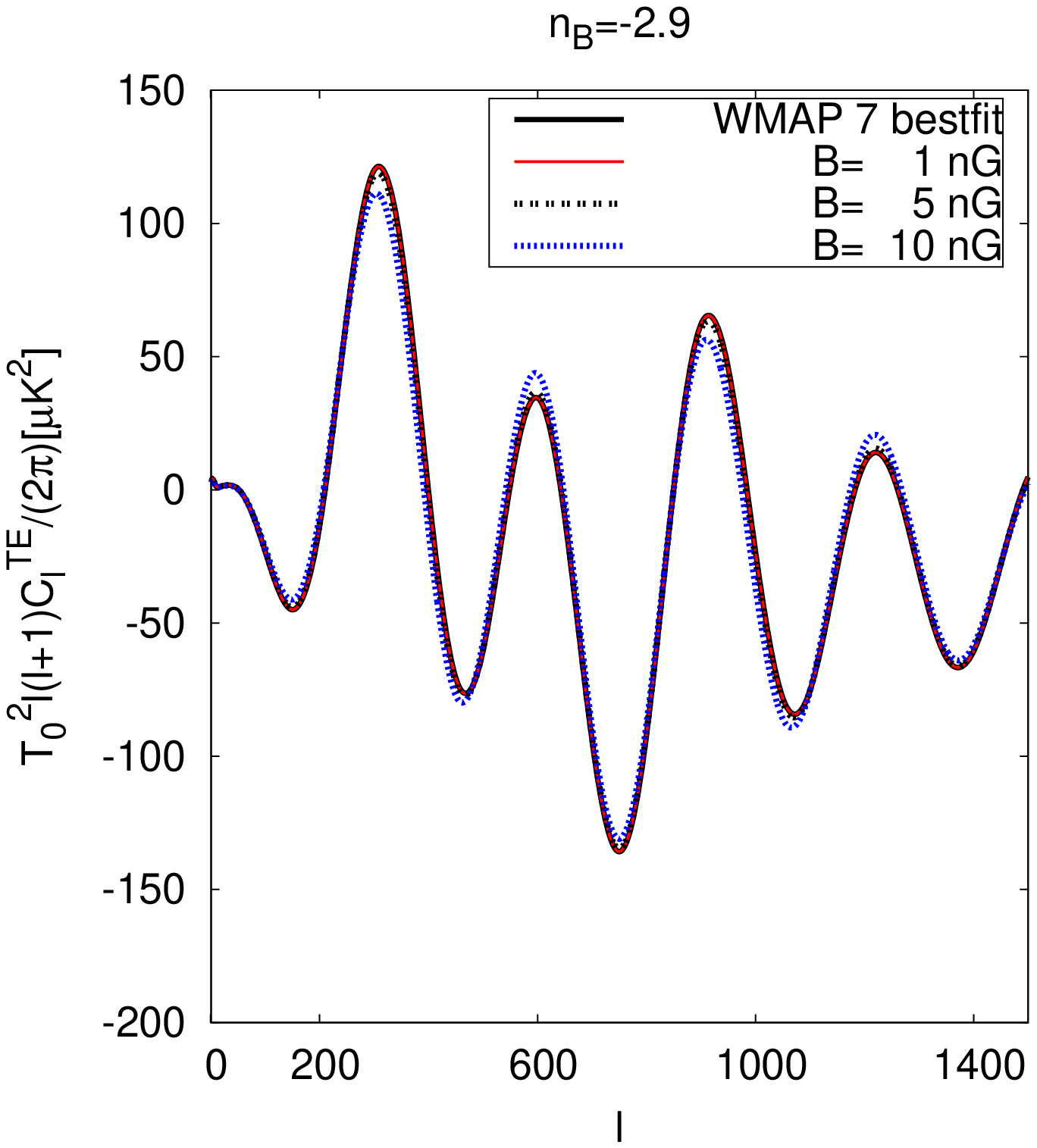 }}
\caption{ The angular power spectra determining the autocorrelation functions of the temperature anisotropies of the CMB ($C_{\ell}^{\rm TT}$), the polarization of the CMB ($C_{\ell}^{\rm EE}$) and the temperature polarization cross correlation function ($C_{\ell}^{\rm TE}$) for different values of the magnetic field strength while keeping fixed the magnetic field spectral index  $n_{\rm B}=-2.9$. These have been calculated using the bestfit values of the six-parameter $\Lambda$CDM fit of WMAP7. For comparison, the pure $\Lambda$CDM model corresponding to $B=0$  has also been shown.}
\label{fig3}
\end{figure}

Using the modified version of CMBEASY including the contributions due to a stochastic magnetic field present before decoupling whose spectrum is effectively cut-off at the diffusion scale using a gaussian window function (cf. equation (\ref{P})) the angular power spectra of the CMB temperature anisotropies and the polarization are calculated.
In particular, for ${\cal P}_{\Delta_{\rm B}}$, ${\cal P}_{\pi_{\rm B}}$ and ${\cal P}_L$ the numerical interpolation as shown in figures \ref{fig1} and \ref{fig2}  is used. This is much more accurate than the approximate solutions given in equations (\ref{appD}), (\ref{appP}) and (\ref{appL}), but at the same time it is much faster than a full numerical integration for each step in $k$-space.
The CMB anisotropies have been calculated by using the initial conditions (\ref{ic})  as a whole including the curvature perturbation and  the part involving the magnetic field. 
This corresponds to a complete correlation between the primordial curvature and the magnetic field \cite{fin1}. 
Moreover, the primordial curvature perturbation is assumed to be determined by the bestfit parameters of the $\Lambda$CDM model of WMAP7 \cite{wmap7}.
In the numerical solution the values for $\Delta_{\rm B}$, $\pi_{\rm B}$ , $L$ and $\zeta$ were determined by the values of the square root of the corresponding power spectrum. In comparison to previous work \cite{GK}  also the power spectrum for the Lorentz term is employed here which is more accurate since $\Delta_{\rm B}$ and $\pi_{\rm B}$ are correlated. This modification manifests itself in  the interesting peak structure in the temperature anisotropies when comparing the cases with and without a magnetic field as discussed below.
In figure \ref{fig3}  the angular power spectra of the temperature autocorrelation function, of the polarization and the temperature polarization cross correlation are shown for different values of the magnetic field parameters. These are compared with the angular power spectra obtained for the  bestfit parameters of the $\Lambda$CDM model of WMAP7 \cite{wmap7}, without a magnetic field. The magnetic field spectral index is chosen to be negative which is the case for magnetic fields generated during inflation (e.g. \cite{tw}) but not for those generated by a causal process, e.g. during the electroweak phase transition (e.g. \cite{ewp}), which requires $n_{\rm B}$ to be an even integer and $n_{\rm B}\geq 2$ \cite{dc}. 

\begin{figure}[h!]
\centerline{\epsfxsize=2.1in\epsfbox{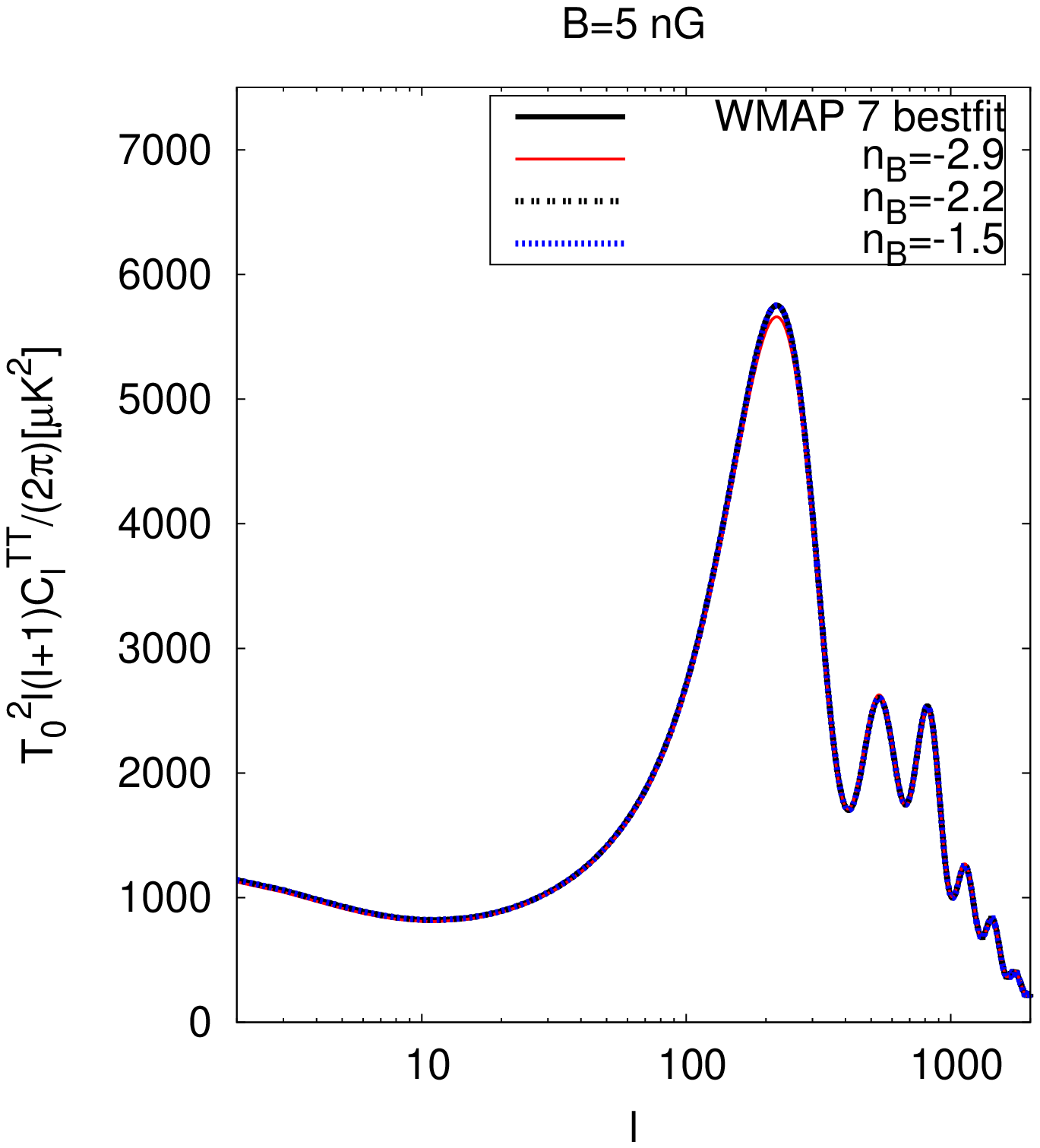}
\hspace{0.2cm}
\epsfxsize=2.1in\epsfbox{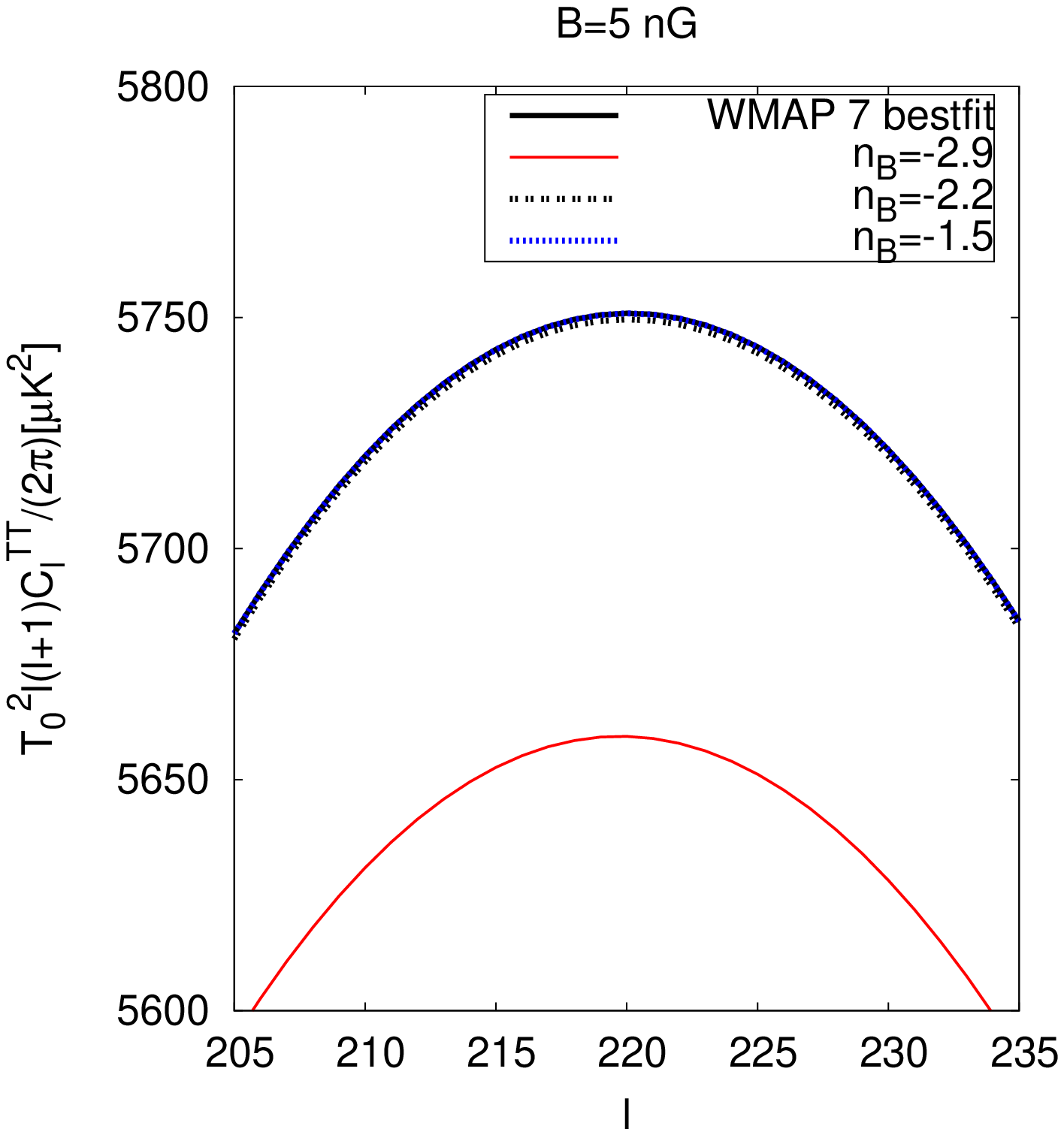}\hspace{0.2cm}
\epsfxsize=2.1in\epsfbox{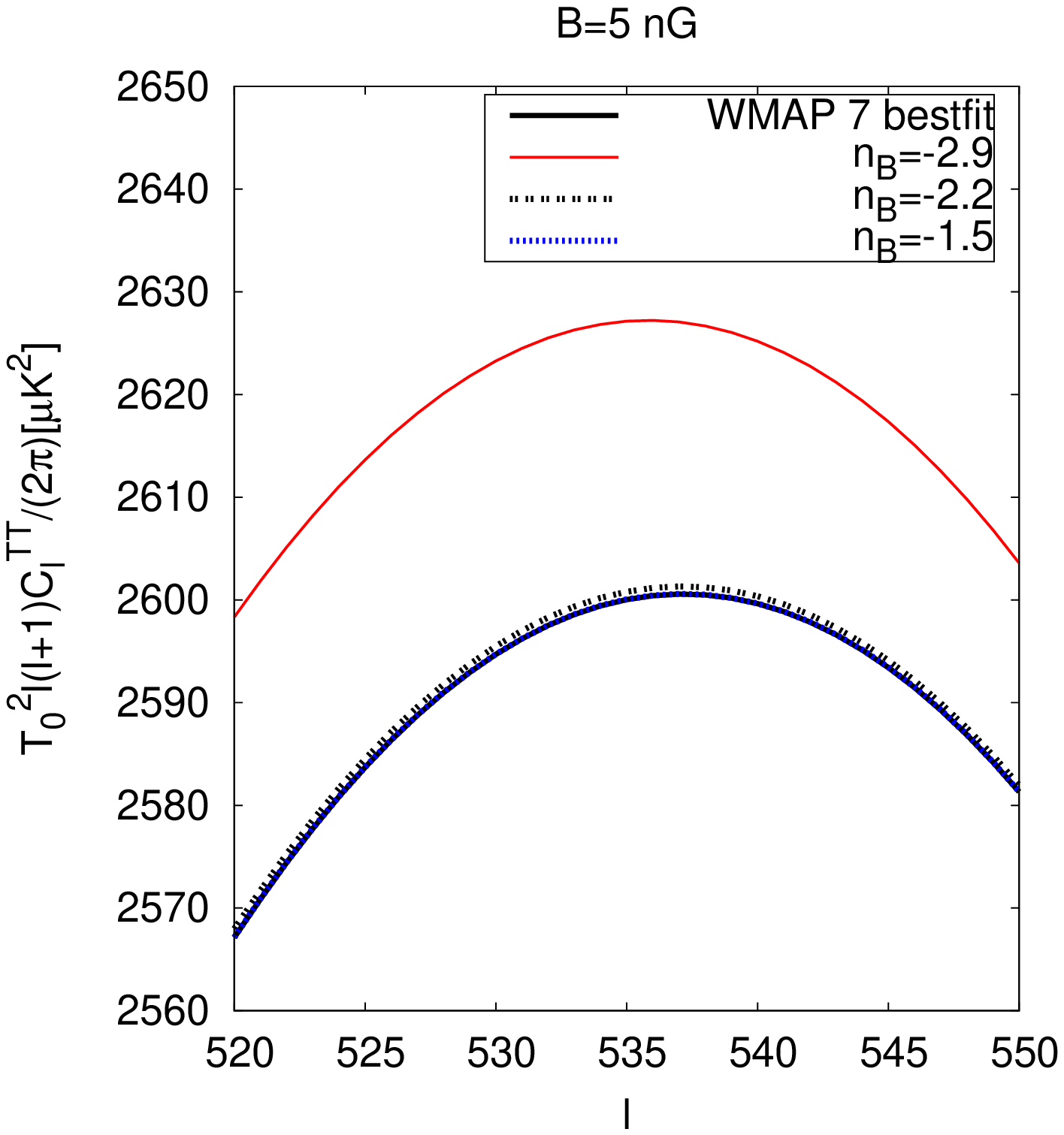 }}
\caption{ The angular power spectra determining the autocorrelation functions of the temperature anisotropies of the CMB ($C_{\ell}^{\rm TT}$) for the magnetic field strength $B= 5$nG while varying the spectral index. 
These have been calculated using the bestfit values of the six-parameter $\Lambda$CDM fit of WMAP7. For comparison, the  $\Lambda$CDM model with  $B=0$  has also been shown.
{\it Left:}  The $TT$ angular power spectrum for the whole range of multipoles. {\it Middle:} The first acoustic peak. {\it Right: } The second acoustic peak.}
\label{fig4}
\end{figure}

\begin{figure}[h!]
\centerline{\epsfxsize=3.in\epsfbox{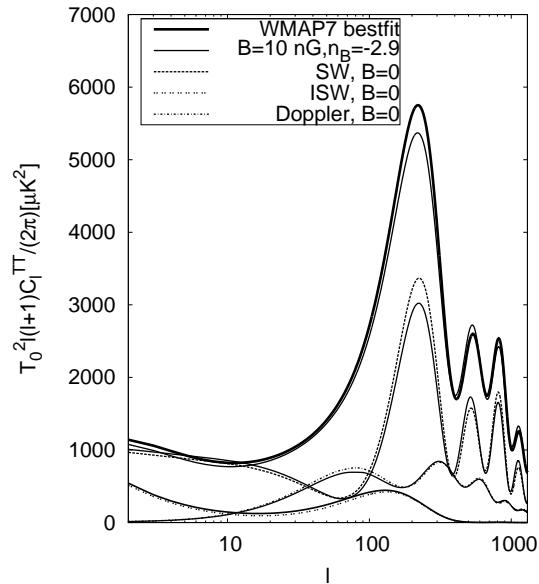}}
\caption{ The different contributions to the total temperature angular power spectrum due to the Sachs-Wolfe term (SW), the integrated Sachs-Wolfe effect (ISW) and the Doppler term. The light solid line shows always the case in the presence of a magnetic field with $B=10$ nG and $n_{\rm B}=-2.9$.}
\label{fig5}
\end{figure}

Figures \ref{fig3} and \ref{fig4} have been calculated using the bestfit values of the six parameter $\Lambda$CDM model of WMAP7 \cite{wmap7},  that is in particular, $\Omega_b=0.0445$, $\Omega_{\Lambda}=0.738$, $\Delta_{\cal R}^2=2.38\times 10^{-9}$, $n_s=0.969$ and the reionization  optical depth $\tau=0.086$. 
In figure \ref{fig5} the contributions from the Sachs-Wolfe effect, the integrated Sachs-Wolfe effect and the Doppler term are shown for a model with a magnetic field and the bestfit $\Lambda$CDM model of WMAP7.

\subsection{The Sachs-Wolfe effect}

On large scales the temperature perturbation, written in the  gauge invariant variables used here, is determined by \cite{GL,BC}
\begin{eqnarray}
\frac{\delta T}{T}=\left[\frac{\Delta_{\gamma}}{4}+\Psi-\Phi\right]_{\tau_{\rm LS}}+\int_{\tau_{\rm LS}}^{\tau_0}\left(\dot{\Psi}-\dot{\Phi}\right)d\lambda,
\label{dt}
\end{eqnarray}
where decoupling of the photons from the baryon fluid takes place instantaneously at the time of last scattering $\tau_{\rm LS}$ and $\tau_0$ corresponds to the present time.
Moreover the Doppler term describing the relative motion between emitter and observer, namely $V_{{\rm b}\;j}n^j$ has been neglected since on large scales scales at last scattering it is negligible. This can also be appreciated in figure \ref{fig5}. The curvature perturbation on slices of uniform total energy density in terms of the gauge-invariant variables used here is given by \cite{dmsw}
\begin{eqnarray}
\zeta=\frac{\sum_{\alpha}\Delta_{\alpha}\Omega_{\alpha}}{\sum_{\alpha}3(1+w_{\alpha})\Omega_{\alpha}},
\end{eqnarray}
which on large scales is equivalent to  equation (\ref{curv}) and also
\begin{eqnarray}
\zeta=\Phi-\frac{2}{3}\frac{\Psi-{\cal H}^{-1}\dot{\Phi}}{1+w},
\end{eqnarray}
which during matter domination on large scales implies the standard result $\zeta=\frac{5}{3}\Phi$ (cf, e.g., \cite{GL}). The total curvature perturbation for the initial  conditions  (\ref{ic}) is given by
\begin{eqnarray}
\zeta=\frac{\Delta_{\gamma}}{4}+\frac{\Omega_{\gamma}\Delta_{\rm B}}{\sum_{\alpha}3(1+w_{\alpha})\Omega_{\alpha}}.
\label{zeta}
\end{eqnarray}
The photon density contrast  $\Delta_{\gamma}$ is approximately constant on large scales (cf. equation (\ref{dg})) so that for the case of no magnetic field the curvature perturbation is conserved on large scales. However, in the presence of the magnetic it can be seen from equation (\ref{zeta}) that $\zeta$ is approximately conserved on large scales during the radiation dominated epoch but not during matter domination where it behaves as $\zeta=c_1+c_2\tau^{-2}$, where $c_1$ and $c_2$ are constants.
Assuming that last scattering takes place in the matter dominated era the curvature perturbation at $\tau_{\rm LS}$ is given by,
\begin{eqnarray}
\zeta(\tau_{\rm LS})=\frac{\Delta_{\gamma}}{4}+\frac{\Omega_{\gamma}(\tau_{\rm LS})}{3}\Delta_{\rm B}.
\end{eqnarray}
Therefore the gauge potential $\Phi$ on large scales is given at the time of last scattering by
\begin{eqnarray}
\Phi(\tau_{\rm LS})=\frac{3}{5}\zeta(\tau_i)-\frac{3}{5}\left[\frac{\Omega_{\gamma}(\tau_i)}{4}-\frac{\Omega_{\gamma}(\tau_{\rm LS})}{3}\right]\Delta_{\rm B},
\end{eqnarray}
where $\tau_i$ refers to the time when the initial conditions are set deep inside the radiation dominated era. Clearly, in the case of no magnetic field $\zeta(\tau_i)=\zeta(\tau_{\rm LS})$ and the standard result is recovered.
Finally,  neglecting the integrated Sachs Wolfe effect (the integral in equation (\ref{dt})) the temperature perturbation is found to be  
\begin{eqnarray}
\frac{\delta T}{T}\simeq -\frac{1}{5}\zeta(\tau_i)+\frac{1}{5}\left[\frac{\Omega_{\gamma}(\tau_i)}{4}-2\Omega_{\gamma}(\tau_{\rm LS})\right]\Delta_B.
\end{eqnarray}
Since $\Omega_{\gamma}(\tau_i)$ is much larger than $\Omega_{\gamma}(\tau_{\rm LS})$ it can be seen that the effect of the magnetic field is to  reduce the temperature fluctuation on large scales. Thus the result of \cite{GK} derived in the synchronous gauge (see also \cite{MG2}) is recovered. This indicates that effectively the Sachs-Wolfe plateau is lower in the presence of the magnetic field than in the case without a magnetic field. This describes qualitatively the behaviour of the temperature fluctuation on large scales when initial conditions are set after neutrino decoupling. The case discussed here is different from the one discussed in \cite{BC} where the evolution since before the magnetic field generation which takes place before neutrino decoupling is taken into account.

\subsection{Acoustic oscillations}

\begin{figure}[h!]
\centerline{\epsfxsize=2.6in\epsfbox{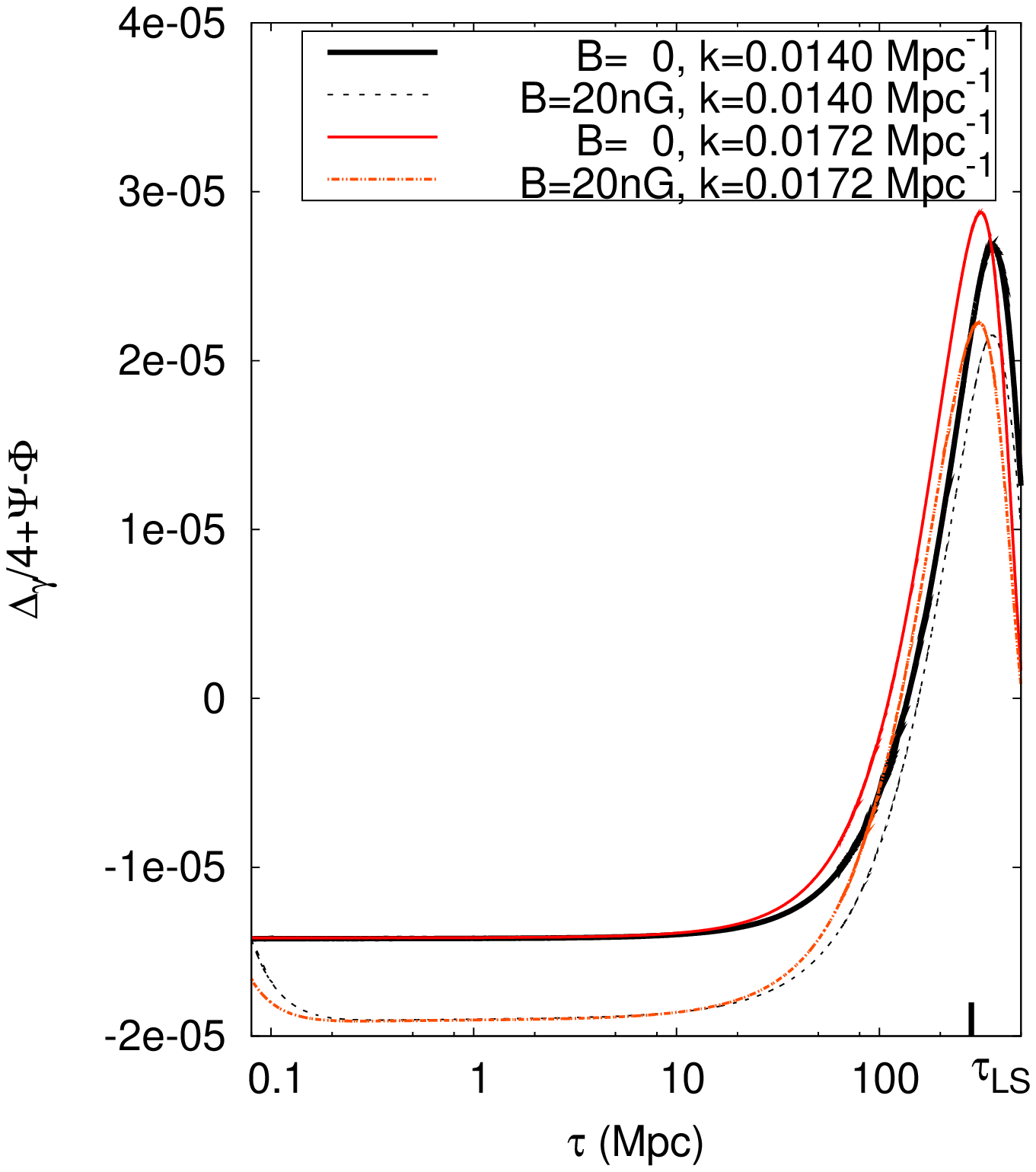}\hspace{0.1cm}
\epsfxsize=2.6in\epsfbox{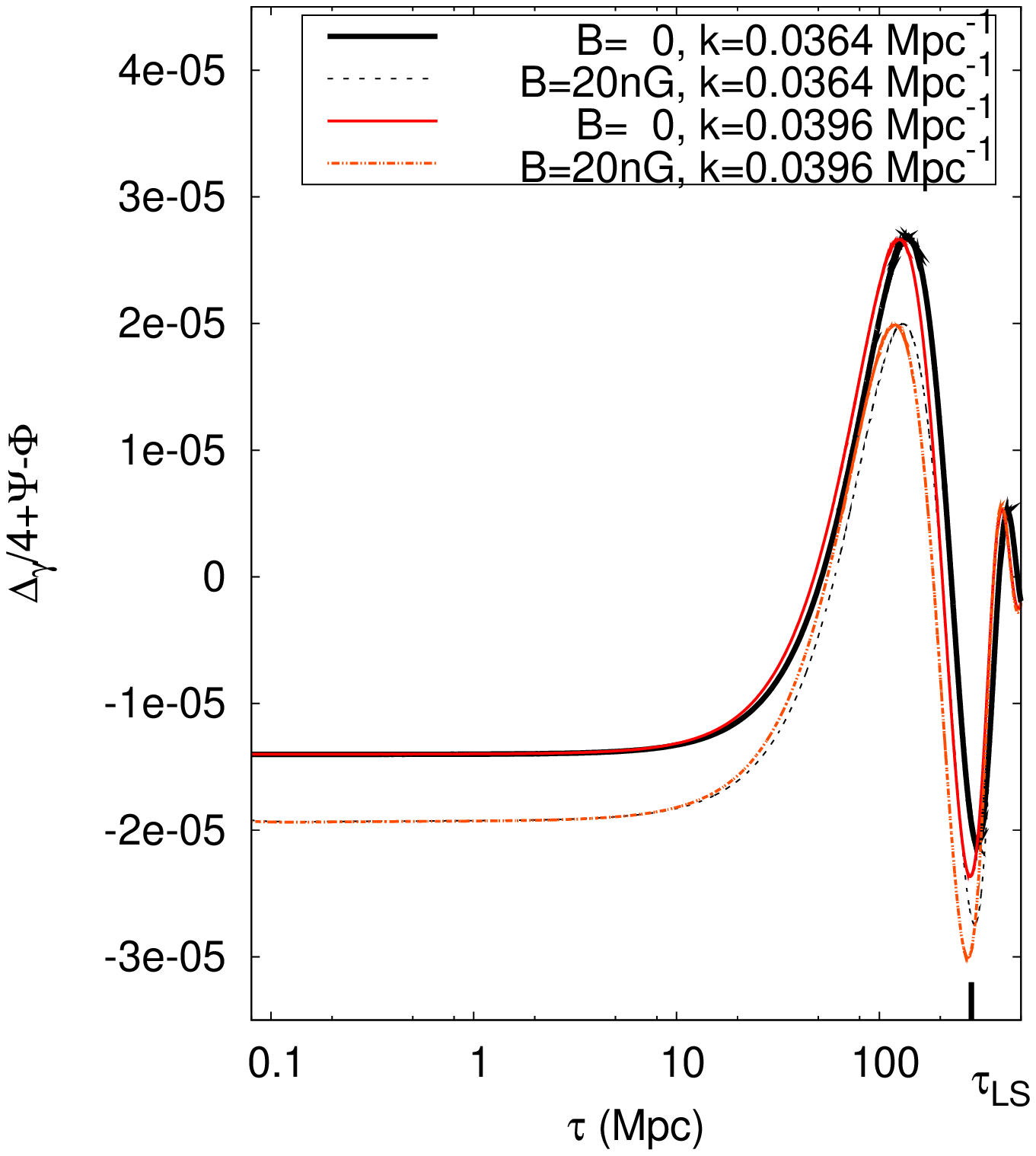 }}
\caption{The numerical solution of the evolution of the effective temperature perturbation is shown for different wave numbers. 
The  WMAP7 bestfit solution is shown in comparison with the solution in presence of a stochastic magnetic field with $B=20$ nG and magnetic spectral index $n_{\rm B}=-2.9$.
Indicated is the time of last scattering, which in this case is $\tau_{\rm LS}=285 $ Mpc.
{\it Left}: The solutions for the wave numbers 
$k=0.0140$ Mpc$^{-1}$ and $k=0.0172$ Mpc$^{-1}$  have a maximum at the time of last scattering.
These wave numbers correspond to the region of the first acoustic peak, whereas the former corresponds to ${\ell=200}$, the latter corresponds to ${\ell=244}$. 
{\it Right}: The solutions for the wave numbers 
$k=0.0364$ Mpc$^{-1}$ and $k=0.0396$ Mpc$^{-1}$  have a minimum at the time of last scattering.
These wave numbers correspond to the region of the second acoustic peak, whereas the former corresponds to ${\ell=523}$, the latter corresponds to ${\ell=569}$.}
\label{fig6}
\end{figure}

The effect of the magnetic field is strongly noticed in the acoustic oscillations of the baryon-photon fluid, as can be appreciated from the Sachs-Wolfe contribution in figure \ref{fig5} (see also figures \ref{fig3} and \ref{fig4}).
These are determined by the photon energy density contrast $\Delta_{\gamma}$.
Using equations (\ref{dg}) and (\ref{vgtc}) its evolution is determined in the tight coupling limit by,
\begin{eqnarray}
\ddot{\Delta}_{\gamma}+\frac{\dot{R}_{\rm b}}{1+R_{\rm b}}\dot{\Delta}_{\gamma}+c_{s\;{\rm b}\gamma}^2k^2\Delta_{\gamma}\simeq -\frac{k^2}{3(1+R_{\rm b})}L+\frac{4k^2}{3}\frac{2+R_{\rm b}}{1+R_{\rm b}}\Phi
\label{Dosc}
\end{eqnarray}
where $R_{\rm b}\equiv\frac{1}{R}$ and $c_{s\;{\rm b}\gamma}^2\equiv\frac{1}{3}\frac{1}{R_{\rm b}+1}$ is the sound speed of the baryon-photon fluid. In general this is solved by, \cite{hs1}
\begin{eqnarray}
\Delta_{\gamma}(\tau)&=&\frac{1}{(1+R_{\rm b})^{\frac{1}{4}}}\left[\Delta_{\gamma}(0)\cos(kr_s(\tau))+\frac{\sqrt{3}}{k}\left[\dot{\Delta}_{\gamma}(0)+\frac{1}{4}\dot{R}_{\rm b}(0)\Delta_{\gamma}(0)\right]\sin(kr_s(\tau))\right.
\nonumber\\
&&
\left.
+\frac{\sqrt{3}}{k}\int_0^{\tau}d\tau'\left(1+R_{\rm b}(\tau')\right)^{\frac{3}{4}}\sin\left[kr_s(\tau)-kr_s(\tau')\right]F(\tau')\right],
\end{eqnarray}
where in the case at hand $F(\tau)\equiv-\frac{k^2}{3(1+R_{\rm b})}L+\frac{4k^2}{3}\frac{2+R_{\rm b}}{1+R_{\rm b}}\Phi$. 
Furthermore, the soundhorizon is defined by $r_s(\tau)\equiv\int_0^{\tau}c_{s\;{\rm b}\gamma}d\tau'$.
Neglecting the changes of $R_{\rm b}$ and $\Phi$ due to the expansion of the universe \cite{hs2}
one arrives at a simple toy model to study the basic effects of the magnetic field.
In this case equation (\ref{Dosc}) reads,
\begin{eqnarray}
(1+R_{\rm b})\ddot{\Delta}_{\gamma}+\frac{k^2}{3}\Delta_{\gamma}=-\frac{k^2}{3}L+\frac{4}{3}k^2(2+R_{\rm b})\Phi, 
\end{eqnarray}
which is solved by
\begin{eqnarray}
\Delta_{\gamma}(\tau)=\left(\Delta_{\gamma}(0)+L-4(2+R_{\rm b})\Phi\right)\cos(\omega\tau)+\frac{\dot{\Delta}_{\gamma}(0)}{\omega}\sin\omega\tau-L+4(2+R_{\rm b})\Phi,
\label{appdpg}
\end{eqnarray}
where $\omega\equiv k/\sqrt{3(1+R_{\rm b})}$.
Here it can be seen that the contribution due to the magnetic field changes as well the amplitude of
the oscillations in the photon energy density contrast as well as the zero point of the oscillations.
Thus the effect of the magnetic field is similar to that of a change in the baryon energy density (cf., e.g.,  \cite{hs1}). Similar to the case of the baryons here  as well an alternating peak structure is found, that is whereas the odd-numbered peaks are lower than in the case without a magnetic field, the even-numbered peaks are larger.
In figure \ref{fig6} numerical solutions for $\frac{\Delta_{\gamma}}{4}+\Psi-\Phi$ 
for several wave numbers are shown. 
The general behaviour can be traced back to the effect the magnetic field contribution has on the oscillations in the photon energy density contrast. 
Going back to equation (\ref{Dosc}) and bearing in mind that $L$ is positive in the model at hand it can be seen that the magnetic field effectively augments the pressure perturbation of the baryon-photon fluid. At last scattering there are mainly two competing effects coming from the gravitational infall determined by the gravitational potential $\Psi$ and the restoring force due to pressure determined in the case at hand  by $c^2_{s\;{\rm b}\gamma}\Delta_{\gamma}$.
The odd-numbered acoustic peaks correspond to the compressional phase of the oscillations \cite{hs1}. Thus increasing the pressure while not changing the gravitational potential significantly results in a larger restoring force and so less effect of the gravitational infall. This reduces the amplitude of odd-numbered peaks in the models with a magnetic field (cf.  figure \ref{fig6}). On the contrary, the even numbered peaks correspond to the expansion phase inside the gravitational well \cite{hs1}. Since the pressure is larger in models with a magnetic field the resulting even-numbered peaks are larger than in the model without a magnetic field. This situation is reflected in figure \ref{fig6} for the second peak which corresponds to a minimum in $\Delta_{\gamma}$ at the time of last scattering. As can be seen in figure \ref{fig6} the minimum in the presence of a magnetic field is deeper than in the case without. Thus leading to a larger amplitude of the second acoustic peak in the presence of a magnetic field. This alternating peak structure is clearly seen in the numerical solutions, cf. figure \ref{fig3}-\ref{fig5}. Moreover, these characteristics are present as well in the extrema of the toy model (\ref{appdpg}). 
Assuming that the initial value of the effective temperature perturbation $\Theta\equiv\frac{\Delta}{4}+\Psi-\Phi$, once it enters the horizon, is of the order of its Sachs-Wolfe amplitude the extrema of $\Theta$ can be estimated.
At $\omega\tau_{\rm LS}=n\pi$, corresponding to a maximum, it is found that $\Theta_{max}=\frac{1}{3}(1+6R_{\rm b})\Phi(\tau_{\rm LS})-\frac{L}{2}$ and at $\omega\tau_{\rm LS}=2n\pi$, corresponding to a minimum $\Theta_{min}=-\frac{1}{3}\Phi(\tau_{\rm LS})$. So that in general the maxima are expected to be smaller in the presence of a magnetic field which results in comparatively smaller values of the odd-numbered acoustic peaks.

\subsection{The linear matter power spectrum}

\begin{figure}[h!]
\centerline{\epsfxsize=3.in\epsfbox{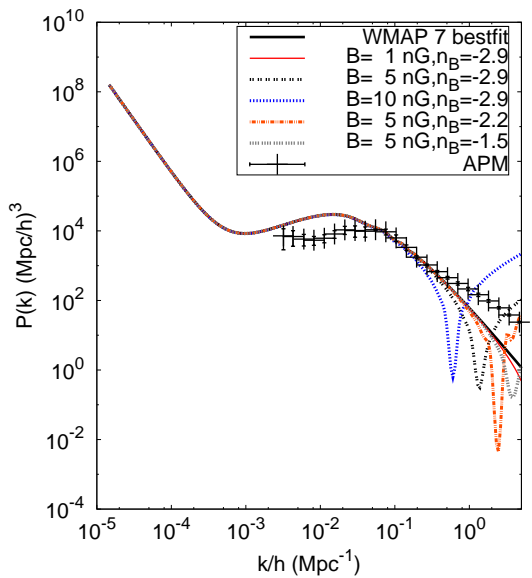}}
\caption{The total matter power spectrum for different spectral index and amplitude of the magnetic field. The data points included show the estimated power spectrum from the APM angular galaxy catalogue \cite{GB} (see also \cite{PGB}).}
\label{fig7}
\end{figure}

During matter domination the total matter perturbation $\Delta_{\rm m}\equiv \tilde{R}_{\rm c}\Delta_{\rm c}+\tilde{R}_{\rm b}\Delta_{\rm b}$, where $\tilde{R}_i$ determines the fractional energy density with respect to the total matter energy density, is determined by,
\begin{eqnarray}
\ddot{\Delta}_{\rm m}+{\cal H}\dot{\Delta}_{\rm m}-\frac{3}{2}{\cal H}^2\Delta_{\rm m}={\cal H}^2\Omega_{\gamma}\Delta_{\rm B}-\frac{k^2}{3}\Omega_{\gamma}L.
\label{dm}
\end{eqnarray}
In deriving equation (\ref{dm}) it has been used that $\Psi\simeq -\Phi$ and that during matter domination on small scales, $\Phi\simeq\frac{a^2\bar{\rho}\Delta}{2\bar{M}_{\rm P}^2k^2}$ which yields to $\Phi\simeq\frac{3{\cal H}^2}{2k^2}(\Delta_{\rm m}+\Omega_{\gamma}\Delta_{\rm B})$.
Thus on very small scales, the contribution from the magnetic field on the righthandside of equation (\ref{dm}) dominates. In this limit there is a constant solution of equation (\ref{dm}), namely,
$\Delta_{\rm m}\propto k^2L$. This implies that for large values of $k$ the matter power spectrum scales as the corresponding power spectrum of the Lorentz term, ${\cal P}_{\Delta_{\rm m}}\propto k^4{\cal P}_{\rm L}$ \cite{sl10}.  
The total matter power spectrum ${\cal P}_{\Delta_{\rm m}}$ is shown in figure \ref{fig7} which clearly shows the influence of the magnetic field on small scales. 
The resulting linear matter power spectra are similar to what is  presented in \cite{fin1}.
However, the general shape  differs from that reported in \cite{sl10} which only considers the compensated magnetic mode without the contribution from the primordial curvature mode and uses a different form for the power spectra related to the magnetic contribution.
The normalization of the matter power spectrum $\sigma_8$ is defined to be the variance of a mass fluctuation inside a sphere of radius $R=8h^{-1}$ Mpc, that is (see for example, \cite{eh})
\begin{eqnarray}
\sigma_8=\int_0^{\infty}\frac{dk}{k}{\cal P}_{\Delta_{\rm m}}\left[\frac{3j_1(kR)}{kR}\right]^2,
\end{eqnarray}
where $j_1(x)=(\sin x-x\cos x)/x^2$. Calculating $\sigma_8$ for the different values of the spectral index and strength of the magnetic field used in the examples above it is found that for $n_{\rm B}=-2.9$ $\sigma_8$ decreases with the strength of the magnetic field, so that for $B=10$ nG it is found that $\sigma_8=0.704$ whereas for the $\Lambda$CDM model without magnetic field $\sigma_8=0.802$. On the other hand varying the spectral index and keeping the magnetic field strength fixed at $B=5$ nG the largest value is found for $n_{\rm B}=-1.5$ that is $\sigma_8=0.802$ and the smallest for $n_{\rm B}=-2.9$ with $\sigma_8=0.77$.

\subsection{Uncorrelated curvature and compensated magnetic modes}

Before closing this section we will comment on the case in which the primordial curvature mode, which is assumed to not include any magnetic contributions, and the compensated magnetic mode are uncorrelated. 
The passive mode is not considered here, since it was also not considered explicitly in the previous subsections where the curvature perturbation was assumed to be determined by the bestfit values of the six parameter $\Lambda$CDM model of WMAP7 \cite{wmap7}.

\begin{figure}[h!]
\centerline{\epsfxsize=3.in\epsfbox{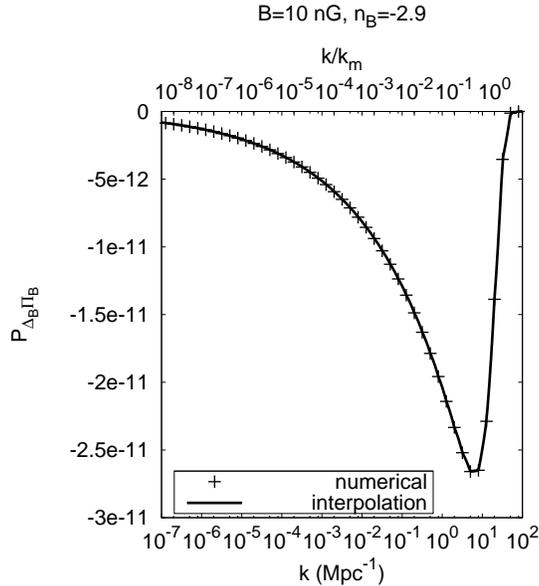}}
\caption{The spectrum determining the crosscorrelation function of the 
magnetic energy density and anisotropic stress is shown  for $B=10$ nG and spectral index $n_{\rm B}=-2.9$ . The numerical solution of the double integral, together with the numerical interpolation (spline) used in the code to calculate the CMB anisotropies is plotted. The lower horizontal axis shows $k$, the upper one shows the ratio $k/k_{\rm m}$. In this example, the magnetic damping wave number is given by  $k_{\rm m}=20$ Mpc$^{-1}$ (cf. equation (\ref{km})). }
\label{fig8}
\end{figure}

\begin{figure}[h!]
\centerline{\epsfxsize=2.1in\epsfbox{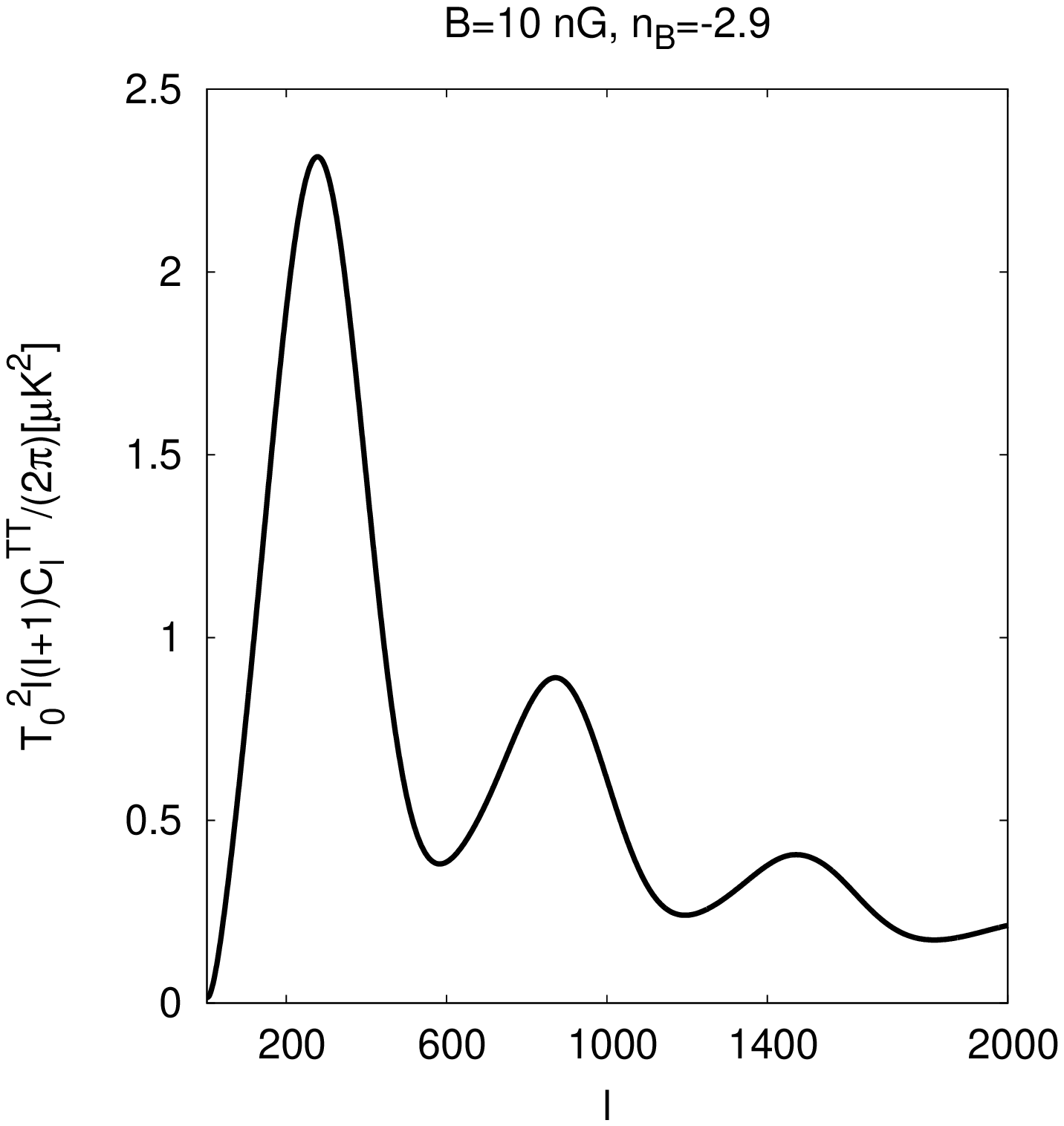}
\hspace{0.05cm}
\epsfxsize=2.1in\epsfbox{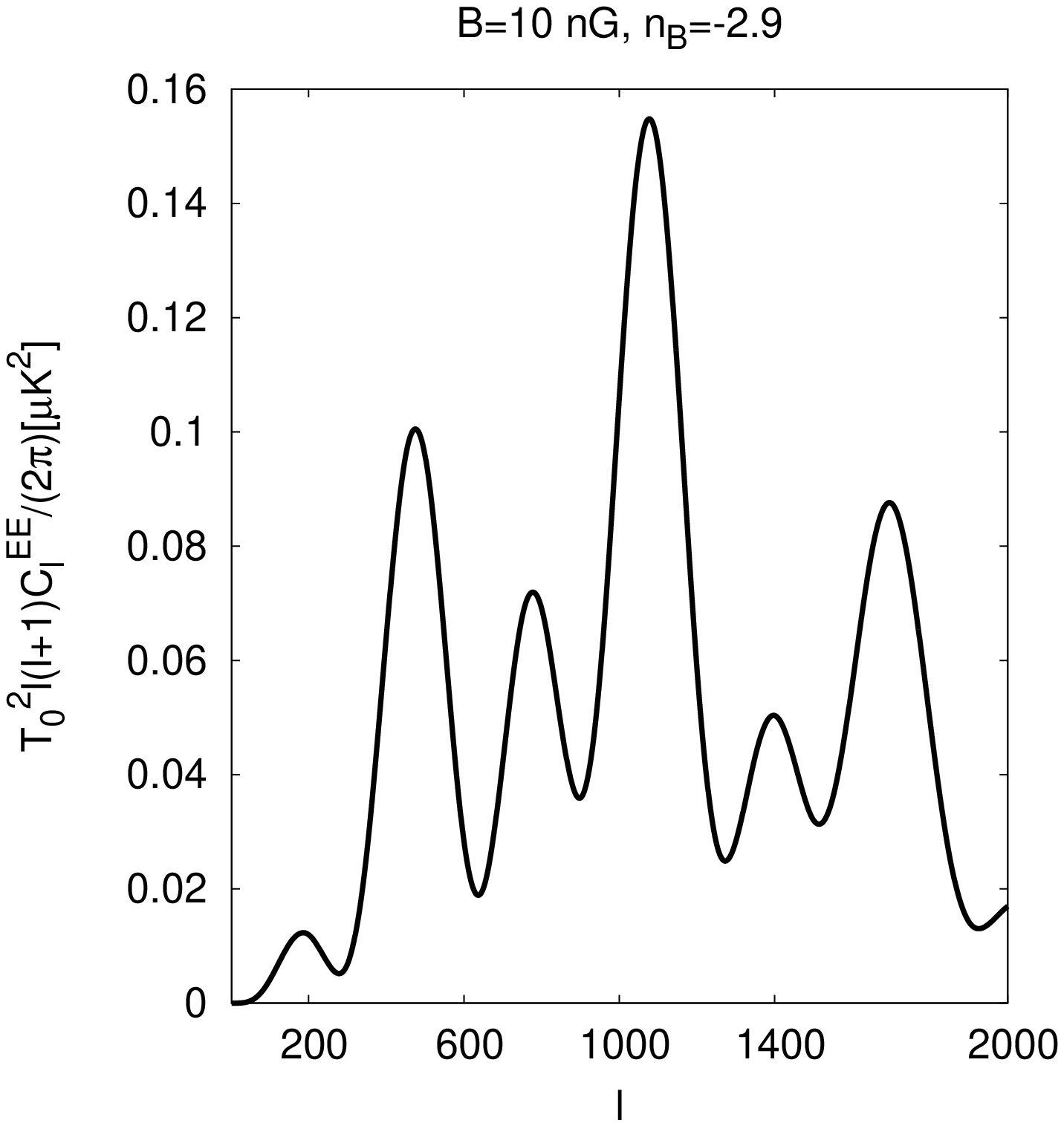}\hspace{0.05cm}
\epsfxsize=2.1in\epsfbox{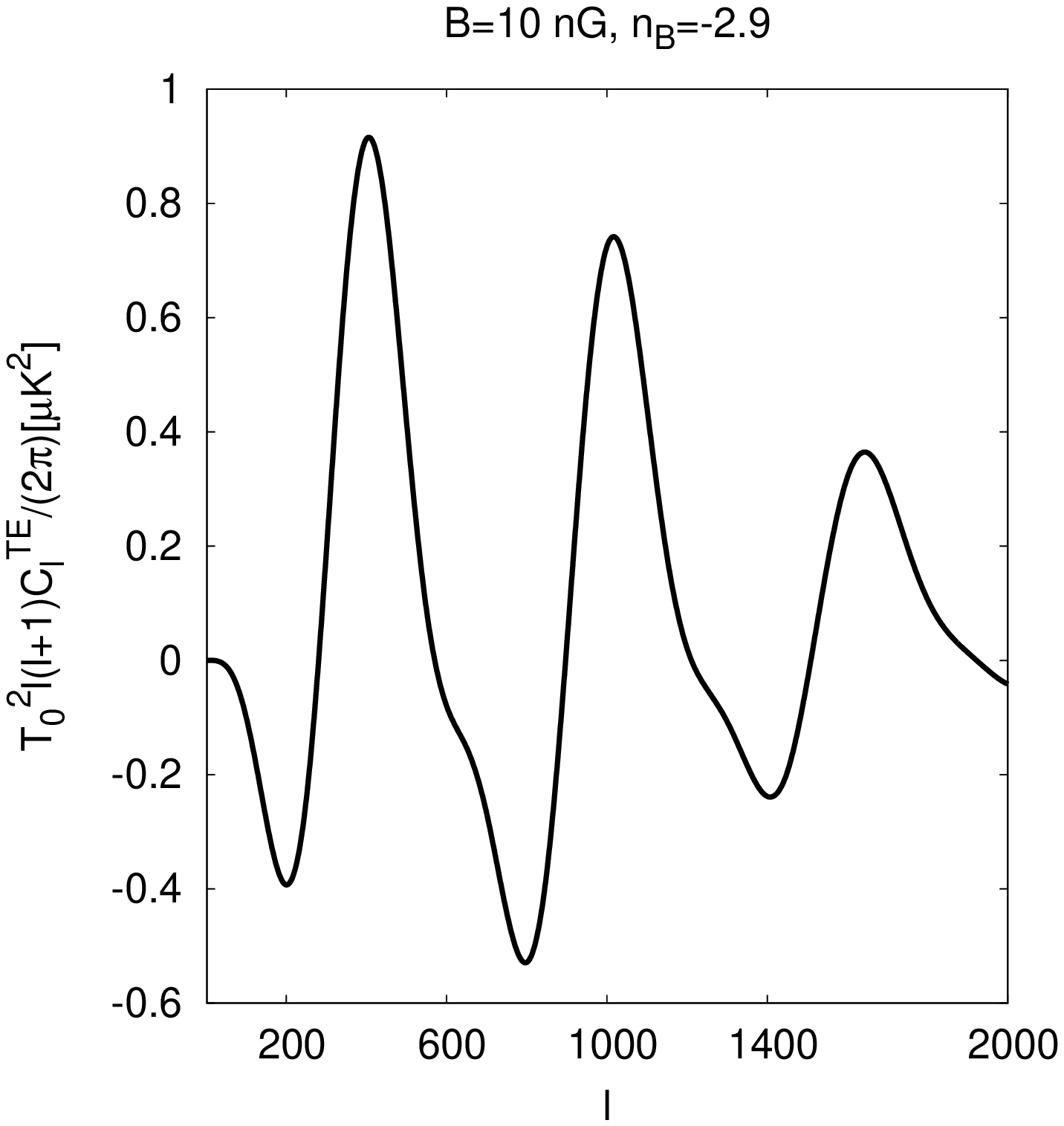 }}
\caption{ The angular power spectra determining the autocorrelation functions of the temperature anisotropies of the CMB ($C_{\ell}^{\rm TT}$) (cf. equation (\ref{ctt})), the polarization of the CMB ($C_{\ell}^{\rm EE}$) and the temperature polarization cross correlation function ($C_{\ell}^{\rm TE}$) (cf. equation (\ref{cte}))  for the compensated  magnetic mode for  $B=10$ nG and $n_{\rm B}=-2.9$. These have been calculated using the bestfit values of the six parameter $\Lambda$CDM fit of WMAP7.}
\label{fig9}
\end{figure}

The compensated magnetic mode corresponds to 
zero inital curvature \cite{le} and we follow the treatment of correlated isocurvature initial conditions (see for example, \cite{iso}).  In this case the initial conditions (cf. equations (\ref{ic})) with vanishing total curvature perturbation ($\zeta=0$) have two different contributions, one proportional to the magnetic energy density $\Delta_{\rm B}$ and one proportional to the magnetic anisotropic stress $\pi_{\rm B}$. 
Moreover, the Lorentz term can also be written in terms of $\Delta_{\rm B}$ and $\pi_{\rm B}$, (cf. equation (\ref{lorentz})). 
Thus the initial conditions as well as the evolution equations only contain terms proportional to  the magnetic energy density or the magnetic anisotropic stress, respectively.
Therefore we define the total brightness function $\hat{\Theta}_{\ell}$ due to the compensated magnetic mode by
\begin{eqnarray}
\hat{\Theta}_{\ell}(\vec{k})=G_{\ell}^{\Delta_{\rm B}}(k)\hat{\Delta}_{\rm B}(\vec{k})+G_{\ell}^{\pi_{\rm B}}(k)\hat{\pi}_{\rm B}(\vec{k}),
\end{eqnarray}
where $G_{\ell}^{\Delta_{\rm B}}(k)$ is the transfer function calculated for $\Delta_{\rm B}=1$ and $\pi_{\rm B}=0$ and $G_{\ell}^{\pi_{\rm B}}(\vec{k})$ is the transfer function obtained for 
$\Delta_{\rm B}=0$ and $\pi_{\rm B}=1$.
Moreover, $\hat{\Delta}_{\rm B}$  denotes the Gaussian random variable corresponding to the magnetic energy density and $\hat{\pi}_{\rm B}$ the Gaussian random variable corresponding to the magnetic anisotropic stress. The autocorrelation functions of these variables are determined by equations (\ref{F}), (\ref{PD}) and (\ref{Ppp}). In addition, the cross correlation function of $\hat{\Delta}_{\rm B}$ and $\hat{\pi}_{\rm B}$ is required
\begin{eqnarray}
\langle \hat{\Delta}_{\rm B}^*(\vec{k})\hat{\pi}_{\rm B}(\vec{k}')\rangle=\frac{2\pi^2}{k^3}{\cal P}_{\Delta_{\rm B}\pi_{\rm B}}(k)\delta_{\vec{k},\vec{k}'}
\end{eqnarray}
where
\begin{eqnarray}
{\cal P}_{\Delta_{\rm B}\pi_{\rm B}}(k)=\frac{3}{\left[\Gamma\left(\frac{n_{\rm B}+3}{2}\right)\right]^2}\left[\frac{\rho_{\rm B 0}}{\rho_{\gamma 0}}\right]^2\left(\frac{k}{k_{\rm m}}\right)^{2(n_{\rm B}+3)}e^{-\left(\frac{k}{k_{\rm m}}\right)^2}
\int_0^{\infty}dz z^{n_{\rm B}+2}e^{-2\left(\frac{k}{k_{\rm m}}\right)^2 z^2}
\nonumber\\
\int_{-1}^1dx e^{2\left(\frac{k}{k_{\rm m}}\right)^2zx}
\left(1-2zx+z^2\right)^{ \frac{n_{\rm B}-2}{2}}\left(-1+z^2+zx-(1+3z^2)x^2+3zx^3\right),
\end{eqnarray}
which is  shown in figure \ref{fig8} for particular values of the magnetic field parameters.
The total angular power spectrum determining the temperature autocorrelation function, $C_{\ell}^{TT}$  is given by,
 \begin{eqnarray}
 C_{\ell}^{TT}=\int\frac{dk}{k}\left[{\cal P}_{\Delta_{\rm B}}\left[G_{\ell}^{\Delta_{\rm B}}(k)\right]^2+2{\cal P}_{\Delta_{\rm B}\pi_{\rm B}}G_{\ell}^{\Delta_{\rm B}}(k)G^{\pi_{\rm B}}_{\ell}(k)+{\cal P}_{\pi_{\rm B}}\left[G_{\ell}^{\pi_{\rm B}}(k)\right]^2\right]
 \label{ctt}
 \end{eqnarray}
and a similar expression for the angular power spectrum
determining the autocorrelation function of the E-mode, $C_{\ell}^{EE}$. The temperature-polarization cross correlation angular  power spectrum is given by,
\begin{eqnarray}
C_{\ell}^{TE}&=&\int\frac{dk}{k}\left[{\cal P}_{\Delta_{\rm B}}G_{\ell}^{\Delta_{\rm B}}(k)H_{\ell}^{\Delta_{\rm B}}(k)+{\cal P}_{\Delta_{\rm B}\pi_{\rm B}}\left[G_{\ell}^{\Delta_{\rm B}}(k)H_{\ell}^{\pi_{\rm B}}(k)+G_{\ell}^{\pi_{\rm B}}(k)H_{\ell}^{\Delta_{\rm B}}(k)\right]
\right.
\nonumber\\
&& \left.\hspace{2cm}
+{\cal P}_{\pi_{\rm B}}G_{\ell}^{\pi_{\rm B}}(k)H_{\ell}^{\pi_{\rm B}}\right],
\label{cte}
\end{eqnarray}
where $H_{\ell}^{\Delta_{\rm B}}$ and $H_{\ell}^{\pi_{\rm B}}$ denote the polarization transfer functions for $\Delta_{\rm B}=1$, $\pi_{\rm B}=0$ and $\Delta_{\rm B}=0$, $\pi_{\rm B}=1$, respectively.
The angular power spectra due to the compensated magnetic mode are shown in figure \ref{fig9}.
Assuming that there is no correlation between the magnetic mode and the primordial curvature 
fluctuation the total angular power spectrum is given by the sum of the angular power spectra. Taking into account the order of magnitude of the bestfit WMAP7  $\Lambda$CDM model (cf. figure \ref{fig3}) and the compensated magnetic mode, as shown in figure \ref{fig9} for $B=10 $ nG and $n_{\rm B}=-2.9$, it is seen that the contribution of the magnetic field to the final angular power spectrum is much smaller than when treating the curvature perturbation and the magnetic mode as fully correlated as in  the previous subsections (cf. figure \ref{fig3}).

\begin{figure}[h!]
\centerline{\epsfxsize=2.5in\epsfbox{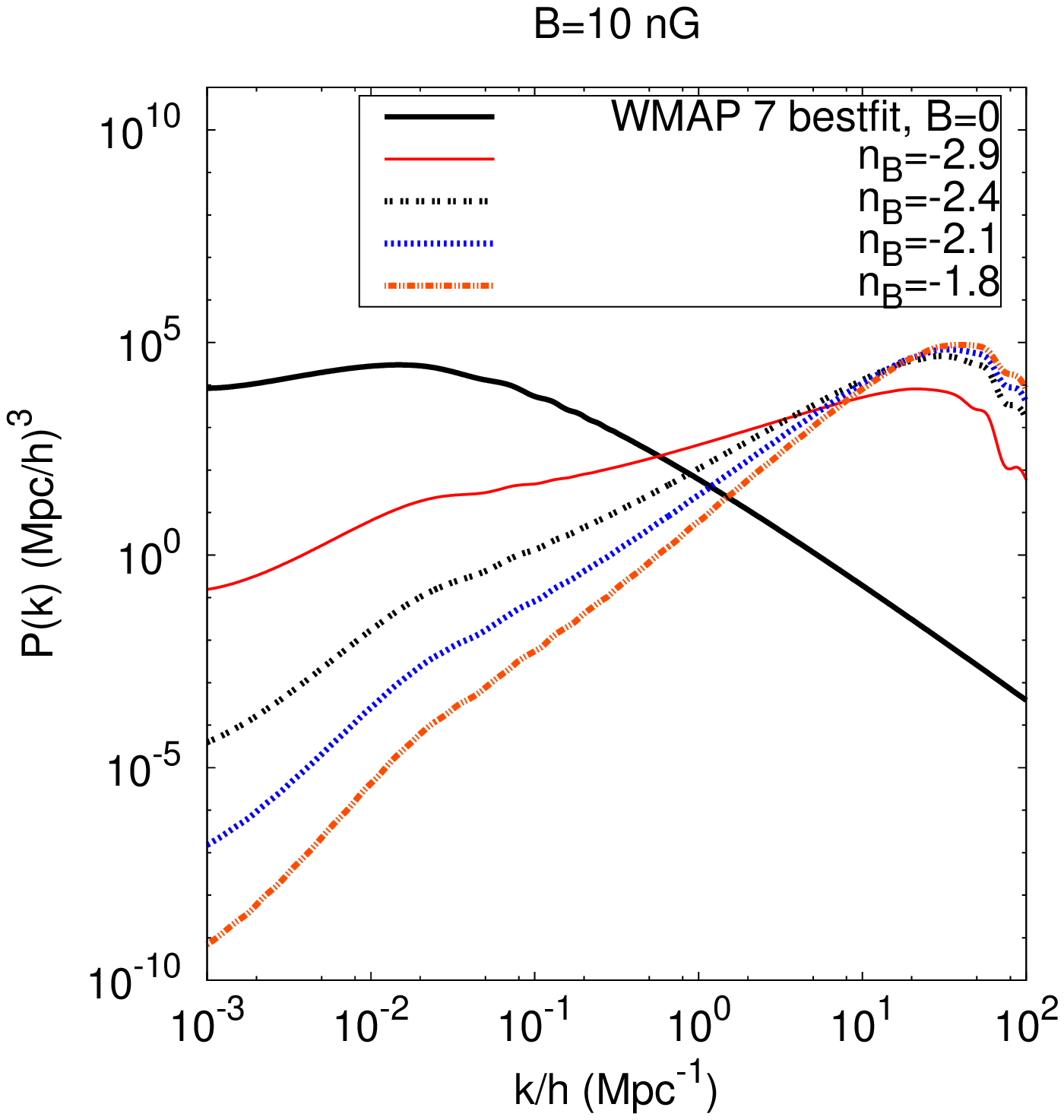}
\hspace{2.5cm}
\epsfxsize=2.5in\epsfbox{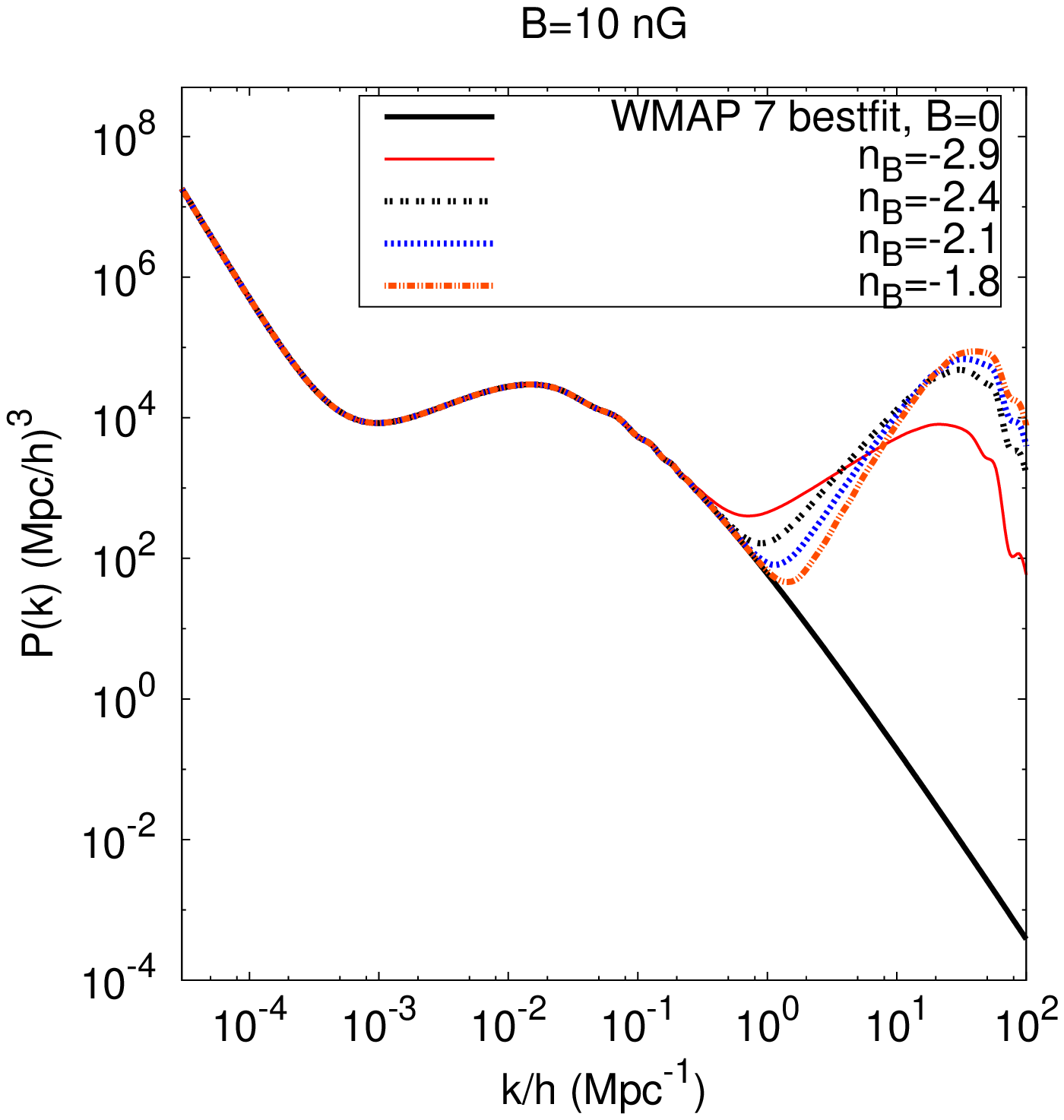}}
\caption{{\it Left}: The total linear matter power spectrum due to the compensated magnetic mode.
For comparison the total linear matter power spectrum for the WMAP7 bestfit model is included
which corresponds to a pure adiabatic mode. 
{\it Right}: The total linear matter power spectrum due to the adiabatic mode and the compensated magnetic mode assuming that these are uncorrelated.
Both graphs  have been calculated using the bestfit values of the six parameter $\Lambda$CDM fit of WMAP7.}
\label{fig10}
\end{figure}

Finally, the total linear matter perturbation due to the compensated magnetic mode  is given by
\begin{eqnarray}
\hat{\Delta}_{\rm m}=\tilde{R}_{\rm b}\left(\hat{\Delta}_{\rm b,\Delta_{\rm B}}+\hat{\Delta}_{\rm b,\pi_{\rm B}}\right)+\tilde{R}_{\rm c}\left(\hat{\Delta}_{\rm c,\Delta_{\rm B}}+\hat{\Delta}_{\rm c,\pi_{\rm B}}\right),
\end{eqnarray}
where $\hat{\Delta}_{i,\Delta_{\rm B}}$ and $\hat{\Delta}_{i,\pi_{\rm B}}$ denote the final density perturbations proportional to $\hat{\Delta}_{\rm B}$ and $\hat{\pi}_{\rm B}$, respectively, and $i$ denotes b or c. In the former case only the contribution due to the magnetic energy density is taken into account in the initial conditions and the evolution equations and in the latter only the one due to the magnetic anisotropic stress.
Therefore, the total linear  matter power spectrum due to the compensated magnetic mode is found to be
\begin{eqnarray}
P_{\Delta_{\rm m}}(k)=\frac{2\pi^2}{k^3}\left[\tilde{R}_{\rm b}^2 U_{\rm b}+2\tilde{R}_b\tilde{R}_c V
+\tilde{R}_{\rm c}^2U_{\rm c}\right],
\end{eqnarray}
where
\begin{eqnarray}
U_{i}&\equiv&{\cal P}_{\Delta_{\rm B}}\Delta_{i,\Delta_{\rm B}}^2+2{\cal P}_{\Delta_{\rm B}\pi_{\rm B}}\Delta_{i,\Delta_{\rm B}}\Delta_{i,\pi_{\rm B}}+{\cal P}_{\pi_{\rm B}}\Delta_{i,\pi}^2
\hspace{2cm}i={\rm b}, \;{\rm c}\nonumber\\
V&\equiv& {\cal P}_{\Delta_{\rm B}}\Delta_{\rm b,\Delta_{\rm B}}\Delta_{\rm c,\Delta_{B}}+{\cal P}_{\Delta_{\rm B}\pi_{\rm B}}\left(\Delta_{\rm b,\Delta_{\rm B}}\Delta_{\rm c,\pi_{\rm B}}+\Delta_{\rm b,\pi_{\rm B}}\Delta_{\rm c,\Delta_{\rm B}}\right)+{\cal P}_{\pi_{\rm B}}\Delta_{\rm b,\pi_{\rm B}}\Delta_{\rm c,\pi_{\rm B}},
\end{eqnarray}
where the definitions $\hat{\Delta}_{i,\Delta_{\rm B}}\equiv \hat{\Delta}_{\rm B}\Delta_{i,\Delta_{\rm B}=1,\pi_{\rm B}=0}$ and $\hat{\Delta}_{i,\pi_{\rm B}}\equiv\hat{\pi}_{\rm B}\Delta_{i,\pi_{\rm B}=1,\Delta_{\rm B}=0}$ were used.
The total matter power spectrum due to the compensated magnetic mode is shown in figure \ref{fig10} ({\it left}) for different values of the magnetic field parameters which is similar to the result of \cite{sl10}, but not identical due to the gaussian window function used here in the calculation of the different magnetic spectra. Assuming that the standard adiabatic mode, due to a primordial curvature perturbation, and the magnetic mode are uncorrelated the total matter power spectrum for the WMAP7 $\Lambda$CDM model is given by the sum of the matter power spectra which is shown in figure \ref{fig10} ({\it right}).
In this case the resulting value of $\sigma_8$ including the contributions from the pure adiabatic mode of the bestfit WMAP7 $\Lambda$CDM model and the compensated magnetic mode is always larger than in the case with no magnetic field.

\section{Conclusions}

The CMB anisotropies in the presence of a stochastic magnetic field have been calculated using the gauge invariant formalism. 
Initial conditions have been found by imposing adiabaticity between all matter/radiation components except the magnetic contribution. This ensures that in the case without a magnetic field the standard adiabatic mode is recovered. 
The numerical solutions were calculated  with a modified version of  the  CMBEASY programme \cite{cmbeasy}. 
The magnetic field is assumed to be a gaussian field whose spectrum contains a gaussian window function effectively cutting off the spectrum at a wave number corresponding to the magnetic damping scale. 
The magnetic field contributions in the evolution equations require the corresponding power spectra of the magnetic energy density, the anisotropic stress, the Lorentz term and the cross correlation between the magnetic energy density and magnetic stress. These involve convolution integrals which only in the limit of scales much larger than the damping scale can be well approximated in terms of special functions.  
Therefore the numerical solutions have been obtained using a numerical interpolation for these integrals. A strategy similar to the one used in the different Boltzmann solver codes and in particular CMBEASY to calculate the  angular power spectrum of the CMB anisotropies.

Firstly, based on previous work \cite{GK}, the initial conditions are used as a whole, including the curvature mode and  the part of the magnetic field. Effectively, this corresponds to a complete correlation between the primordial curvature perturbation and the magnetic field.
The curvature mode is assumed to be determined by the bestfit values of the six parameter $\Lambda$CDM model of WMAP7 \cite{wmap7}.
The magnetic field variables are determined by the square root of the corresponding power spectra. The difference to the numerical solutions presented  in \cite{GK} is that here the power spectrum for the Lorentz term was used which is more accurate since the magnetic energy density and anisotropic stress are correlated. Moreover, no approximations have been used for the power spectra related to the magnetic field contributions.
Having obtained the numerical solutions for different values of the magnetic field amplitude and spectral index  the results were compared with qualitative estimates on very large scales and on small scales.  In particular the Sachs-Wolfe effect was calculated which was found to be smaller in the presence of a magnetic field which is also what is observed in the numerical solutions. 
Moreover the acoustic oscillations in the photon-baryon fluid were described in a simple model and the alternating peak structure recovered which is seen in the numerical solution for the type of magnetic field spectra used here.

Secondly, the case of uncorrelated curvature and compensated magnetic modes is considered.
The  compensated magnetic mode is considered to be a correlated isocurvature mode consisting of two modes, one proportional to the magnetic energy density and one proportional to the magnetic anisotropic stress. The final angular power spectra determining the temperature and polarization auto- and cross-correlations are calculated taking into account  the cross correlation between the magnetic energy density and anisotropic stress. Assuming no correlation of the compensated magnetic mode with the standard adiabatic mode due to a primordial curvature fluctuation the overall effect of the magnetic field contribution is much smaller than for the same magnetic field values 
in the fully correlated case. Moreover, the total angular power spectra resulting from the standard adiabatic mode and the compensated magnetic mode are always larger in the case including the magnetic field.

Finally, the linear matter power spectrum was calculated  in both cases. At small scales it is observed that it is dominated by the magnetic field contribution which is also seen in the second order differential equation determining the total matter density contrast derived in this limit. The total linear  matter power spectrum is found to have distinct shapes in the two different cases assuming correlated or uncorrelated curvature and magnetic modes, respectively. 
Moreover, it is also noted that the normalization of the matter density fluctuation $\sigma_8$ depends on the magnetic field parameters. This provides yet another possibility of putting limits on the magnetic field parameters \cite{Y}.

\section{Acknowledgements}
 Financial support by Spanish Science Ministry grants FPA2009-10612, FIS2009-07238 and CSD2007-00042 is gratefully acknowledged.

\section{Appendix: Transformation of initial conditions to the synchronous gauge}

The initial conditions (cf. equations (\ref{ic})) are given in terms of the gauge-invariant variables.
The corresponding initial conditions in the synchronous gauge are found by using the 
following.
In general the perturbed metric  is given by \cite{dmsw,ks},
\begin{eqnarray}
ds^2=a^2(\tau)\left[-(1+2A)d\tau^2-2B_id\tau dx^i+(\delta_{\ij}+2H_{ij})dx^idx^j\right]
\end{eqnarray}
where  the expansion in harmonic functions $Y(\vec{k},\vec{x})$, as defined in section 3, for scalar perturbations leads to
\begin{eqnarray}
B_i=BY_i\hspace{2cm}
H_{ij}=H_{\rm L}Y\delta_{ij}+H_{\rm T}Y_{ij}.
\end{eqnarray}
In the synchronous gauge $A=0$, $B=0$. In this case, it is common to write the metric perturbation  as \cite{mb},
\begin{eqnarray}
h_{ij}(\vec{x},\tau)=\int d^3k  e^{i\vec{k}\cdot\vec{x}}\left[\hat{k}_i\hat{k}_jh(\vec{k},\tau)+(\hat{k}_i\vec{k}_j-\frac{1}{3}\delta_{ij})6\eta(\vec{k},\tau)\right],
\hspace{2cm}
\hat{k}_i\equiv\frac{k_i}{k}, 
\end{eqnarray}
which implies $\eta=-\left(H_{\rm L}+\frac{1}{3}H_{\rm T}\right)$ and $h=6H_{\rm L}$.
The gauge invariant Bardeen potentials $\Psi$ and $\Phi$ are given, respectively,  by \cite{dmsw,ks}, 
$\Psi=A-{\cal H}k^{-1}\sigma-k^{-1}\dot{\sigma}$ and
$\Phi=H_{\rm L}+\frac{1}{3}H_{\rm T}-{\cal H}k^{-1}\sigma$, where $\sigma\equiv k^{-1}\dot{H}_{\rm T}-B$.
Moreover, the gauge-invariant velocity is defined as $V=v-k^{-1}\dot{H}_{\rm T}$. Thus the velocity in the synchronous gauge is found to be,
\begin{eqnarray}
v^S=V-\frac{k}{a(\tau)}\int d\tau' a(\tau')\Psi(\tau').
\end{eqnarray}
Therefore together with equation (\ref{delta}) the initial conditions (\ref{ic}) in the synchronous gauge are given by,
\begin{eqnarray}
\delta_{\gamma}^S&=&\delta_{\nu}^S=\frac{4}{3}\delta_{\rm b}^S=\frac{4}{3}\delta_{\rm c}^S=
-\frac{\Omega_{\gamma}}{2}\left(\Delta_{\rm B}+L+\frac{2}{3}\pi_{\rm B}\right)=-\Omega_{\gamma}\Delta_{\rm B},\nonumber\\
\tilde{v}_{\gamma}^S&=&\tilde{v}_{\rm b}^S=-\frac{\Omega_{\gamma}}{8}\left(\Delta_{\rm B}-\frac{2}{3}\pi_{\rm B}\right)+\frac{1+\Omega_{\nu}}{8}L=\frac{\Omega_{\nu}}{4}\Delta_{\rm B}-\frac{\pi_{\rm B}}{6},
\nonumber\\
\tilde{v}_{\nu}^S&=&-\frac{\Omega_{\gamma}}{8}\left(\Delta_{\rm B}+L-\frac{2}{3}\frac{\Omega_{\gamma}+1}{\Omega_{\nu}}\pi_{\rm B}\right)=-\frac{\Omega_{\gamma}}{4}\Delta_{\rm B}+\frac{\Omega_{\gamma}}{\Omega_{\nu}}\frac{\pi_{\rm B}}{6},
\nonumber\\
\tilde{v}_{\rm c}^S&=&0\nonumber\\
\tilde{\pi}_{\nu}&=&-\frac{\Omega_{\gamma}}{\Omega_{\nu}}\tilde{\pi}_{\rm B}-\frac{4\zeta}{15+4\Omega_{\nu}}+\frac{\Omega_{\gamma}}{2\Omega_{\nu}}\frac{4\pi_{\rm B}-6\Omega_{\nu}\Delta_{\rm B}}{15+4\Omega_{\nu}}\nonumber\\
\eta&=&-\zeta+\frac{\Omega_{\gamma}}{8}\left(-\Delta_{\rm B}+L+\frac{2}{3}\pi_{\rm B}\right)=-\zeta,
\end{eqnarray}
where $\tilde{v}^S\equiv v^S/x$.
There are two integration constants one in the calculation of $v^S$ and one in the calculation of $h$. These have been chosen such that $v_{\rm c}^S=0$ and there is no constant mode in $h$.
The final expression on the far right of each equation is  obtained using $L=\Delta_{\rm B}-\frac{2}{3}\pi_{\rm B}$. For $\tilde{\pi}_{\rm B}$ only the expression using this relation for $L$ is given.
At this order of $x$, $h=0$. These are the initial conditions to lowest order in $x$ in the synchronous gauge as given in \cite{MG1,GK,fin1}.

\end{document}